\documentclass{article}

\usepackage{amsmath}
\usepackage{amssymb}
\usepackage{braket}
\usepackage{booktabs}
\usepackage{geometry}
\usepackage{graphicx}
\usepackage{multirow}
\usepackage{subfigure}
\usepackage{threeparttable}
\usepackage{xcolor}

\definecolor{eye-caring}{RGB}{205,222,194}

\title{Enhancing Variational Quantum Circuit Training: An Improved Neural Network Approach for Barren Plateau Mitigation}
\author{Zhehao Yi$^{1}$, Yanying Liang$^{1}$, Haozhen Situ$^{1,}$\thanks{Email: situhaozhen@gmail.com (corresponding author)}
\\
{\small $^{1}$College of Mathematics and Informatics, South China Agricultural University, Guangzhou 510642, China}\\
}

\date{}

\begin{document}
\maketitle
\begin{abstract}
Combining classical optimization with parameterized quantum circuit evaluation, variational quantum algorithms (VQAs) are among the most promising algorithms in near-term quantum computing. Similar to neural networks (NNs), VQAs iteratively update circuit parameters to optimize a cost function. However, the training of variational quantum circuits (VQCs) is susceptible to a phenomenon known as barren plateaus (BPs). Various methods have been proposed to mitigate this issue, such as using neural networks to generate VQC parameters.
In this paper, we improve the NN-based BP mitigation approach by refining the neural network architecture and extend its applicability to a more generalized scenario that includes random quantum inputs and VQC structures. We evaluate the effectiveness of this approach by comparing the convergence speed before and after it is utilized. Furthermore, we give an explanation for the effectiveness of this method by utilizing a loss landscape visualization technique and the expressibility metric of VQC. The smoothness of the loss landscape offers an intuitive insight into the method's utility, while the reduction in expressibility accounts for the enhanced trainability. Our research highlights the universal applicability of the NN-based BP mitigation approach, underscoring its potential to drive progress in the development of VQAs across diverse domains.
\end{abstract}

\section{Introduction}
\label{Sect:intro}
Quantum computing has made remarkable strides, propelled by years of dedicated research and development. Unlike traditional computers, which operate using classical bits, quantum computers utilize quantum bits (qubits) that harness phenomena like superposition and entanglement \cite{book}. These distinctive properties enable quantum computers to tackle complex computation and process information in ways that classical systems cannot match. Consequently, quantum computing holds the promise to revolutionize various industries, offering significant potential for transformative applications.

Variational quantum algorithms (VQAs) \cite{vqa} are hybrid algorithms tailored for the noisy intermediate-scale quantum (NISQ) era. These algorithms leverage the strengths of both classical and quantum computing to address a variety of complex problems. VQAs are composed of two main components: a quantum component and a classical component. In the quantum component, data is encoded into a quantum state and processed through a variational quantum circuit (VQC). The resulting quantum state is then measured, and the outcomes are used as inputs for a cost function or other specific calculations within the VQA framework. The classical component involves adjusting the parameters of the VQC based on the quantum computation outcomes, aiming to minimize the cost function. The optimization of VQC parameters is typically performed using gradient descent techniques.

VQAs have made significant advancements in recent years. Among these, quantum approximate optimization algorithms (QAOA) are a typical class of VQAs designed to find approximate solutions for combinatorial optimization problems, often yielding near-optimal solutions within reasonable computational time \cite{ni2024multilevel}. Another typical class of VQAs is variational quantum eigensolvers (VQE), which are designed to approximate the ground state energy of quantum systems \cite{zhang2022variational}.
Variational quantum algorithms for approximately solving semidefinite programs have been proposed \cite{patel2024variational}.
A quantum counterpart to capsule networks has also been introduced, offering potential explainability in quantum machine learning \cite{liu2022quantum}. A meta-learning VQA has been developed, utilizing classical recurrent units to assist quantum computing, effectively learning to find approximate optima \cite{huang2022learning}. Quantum generative adversarial networks (QGANs), which can consist of either quantum or classical generators and discriminators, have been employed to generate both quantum and classical data \cite{situ2020quantum}.
The quantum kernel self-attention network (QKSAN) has been developed, which integrates the data representation strengths of quantum kernel methods with the efficient information extraction capabilities of the self-attention mechanism \cite{zhao2024qksan}.
Additionally, secure delegated VQAs have been proposed, leveraging quantum homomorphic encryption to allow users with limited quantum capabilities to delegate VQA tasks to remote quantum servers while maintaining the privacy of their training data \cite{li2024secure}.
Notably, QGANs \cite{huang2021quantum,huang2021experimental} and deep quantum neural networks \cite{pan2023deep} have been successfully trained on superconducting processors,
while variational quantum classifiers with active learning \cite{ding2023active} have been implemented on a programmable photonic quantum processor.

Despite the significant advancements in VQAs, a critical trainability issue known as barren plateaus (BPs) persists \cite{bp}. When BPs occur, the cost function landscape becomes flat, causing the gradient to vanish and making it difficult to converge to the global optimum. Specifically, the gradient decreases exponentially as the number of qubits increases, meaning that resolving the gradient to a specific accuracy requires an exponentially larger number of measurements. BPs arise due to various factors, including the choice of cost function \cite{cfd}, entanglement \cite{edbp, eibp}, and the presence of noise \cite{nbp}. Although VQCs with high expressibility are theoretically more likely to reach the desired solution, their flatter cost landscapes make them harder to train \cite{erb}. Even gradient-free optimizers struggle with BPs because they rely on cost function differences, which are exponentially suppressed in these regions \cite{ebp}. To mitigate BPs, several strategies have been proposed, such as meta-learning initialization parameters \cite{llqnncnn}, initializing circuits as sequences of identity blocks \cite{blockinit}, and layer-by-layer training \cite{llqnn}.

A neural network (NN) approach to mitigate BPs was proposed, where the parameters of a VQC are generated by a neural network \cite{abp}. This method demonstrated faster convergence, indicating that neural networks can indeed help alleviate BPs. However, some ingredients of VQCs, including the classical input of the data encoding module and the structure of the VQC, are specific in the previous study. It is, therefore, natural to investigate the effectiveness of this NN-based BP mitigation approach in a broader context. In addition to classical data processing, VQCs are also well-suited for quantum data processing, necessitating the consideration of quantum inputs. Recent advancements in VQCs have introduced variable structures, outperforming manually designed, fixed structures \cite{lu2021markovian,he2022quantum,wu2023quantumdarts,he2024gradient,he2024training}. Consequently, focusing solely on a regular VQC structure is insufficient.

In this paper, we extend the VQC input to random quantum states sampled from the uniform (Haar) measure and generalize the VQC structure to random circuits composed of CNOT and single-qubit rotation gates.
We also strengthen the NN-based BP mitigation approach through refinement of the neural network structure.
We compare the convergence speed of the standard quantum circuit (SQC) model with the neural enhanced quantum circuit (NEQC) model. Additionally, we use a loss landscape visualization technique to analyze the smoothness of the loss landscape between these models, offering an intuitive understanding of the method's utility. Furthermore, we examine the changes in expressibility between SQC and NEQC, aligning these findings with the known relationship between expressibility and trainability. This analysis clarifies the rationale behind the NN-based BP mitigation approach.

The remainder of this paper is structured as follows. In section \ref{Sect:background}, we provide an overview of fundamental concepts relevant to the study, along with a review of related works in the field. Section \ref{Sect:method} presents a comprehensive description of the methods employed in our research, including the rationality and construction of random quantum inputs and circuit structures, the concatenation of neural network and VQC, and the metrics and visualization technique we used to conduct the analysis. In section \ref{Sect:experiment}, we delve into the numerical experiment results, offering a detailed discussion of the results and their implications. Finally, section \ref{Sect:conclusion} summarizes the key findings of this study.

\section{Background and related work}
\label{Sect:background}

\subsection{Expressibility estimation}\label{sect:express}

Sim et al. introduced a descriptor known as expressibility, which quantifies how effectively a VQC can generate quantum states that are well representative of the Hilbert space \cite{expressibility}. Based on the Kullback-Leibler (KL) divergence, the expressibility of a VQC $\mathcal{U}(\theta)$ is given by
\begin{align}\label{Eq: expressibility}
    Expr = D_{KL}(P_\mathcal{U}(F) || P_{Haar}(F)) = \sum_{F}P_\mathcal{U}(F)ln{\frac{P_\mathcal{U}(F)}{P_{Haar}(F)}}.
\end{align}
Expressibility is statistically estimated through three main steps: (1) Generate $2K$ quantum states by initializing the VQC with $|0\rangle$ as the input and random parameters $\theta$ drawn uniformly from $[0, 2\pi]$. These $2K$ states are then grouped into $K$ pairs. (2) For each pair of states, calculate the fidelity,  $F=|\langle \phi_1|\phi_2\rangle|^2$, resulting in $K$ fidelity values. (3) Determine the KL divergence between the probability distribution $P_\mathcal{U}(F)$ and $P_{Haar}(F)$, where $P_\mathcal{U}(F)$ is the distribution obtained from the histogram of the $K$ fidelity values,  and $P_{Haar}(F)$ is the distribution of fidelities of two random quantum states sampled from the Haar measure. $P_{Haar}(F)$ is analytically given by $(2^{N}-1)(1-F)^{2^N-2}$, where $N$ is the number of qubits.

Note that a smaller $Expr$ value indicates greater expressibility. According to Ref. \cite{erb}, there is a well-established relationship between expressibility and BPs: higher expressibility generally increases  susceptibility to BPs, while lower expressibility tends to reduce this susceptibility.

\subsection{Barren plateaus}
BPs significantly hinder the trainability of VQCs. BPs refer to regions in the loss landscape where the gradient becomes extremely small, making it difficult for optimization algorithms to converge to the global optimum. Meanwhile, as the gradient approaches zero, the number of measurements required to estimate it accurately grows exponentially. This challenge can cause the training process to slow down or stall, drastically reducing the efficiency of VQC training.

Cerezo et al. proved that when a cost function involves global observables, it results in exponentially vanishing gradients, even with shallow circuits \cite{cfd}. In contrast, a cost function based on local observables yields at worst a polynomially vanishing gradient, provided the circuit depth is $O(\log N)$. These proofs assume that the VQC is structured as an alternating layered ansatz composed of blocks that form local 2-designs.
Patti et al. explored how random entanglement, generated through state evolution with random unitaries, leads to barren plateau formation \cite{edbp}. Additionally, Wang et al. proved that the gradient vanishes exponentially in the number of qubits if the ansatz depth grows linearly with the number of qubits, under the influence of local Pauli noise \cite{nbp}.

Some methods have been proposed to alleviate the influence of BPs. One approach, outlined in Ref. \cite{llqnn}, is layer-by-layer training, which consists of two phases. In the first phase, the ansatz is bulit incrementally by adding layers. The parameters of the newly added layers are initialized to zero to avoid disrupting the current solution. These parameters are updated while those of the previous layers remain fixed. In the second phase, the pre-trained circuit acquired in phase one is further optimized by training larger contiguous partitions of layers at a time. Another effective technique is meta-learning, as discussed in Ref. \cite{llqnncnn}, which involves initializing the parameters based on knowledge gained from related tasks. This approach can enhance the overall training process and reduce the impact of BPs.

\subsection{Using neural network to generate VQC's parameters}

Friedrich and Maziero \cite{abp} proposed using a neural network to generate the parameters of a VQC, demonstrating that this approach can mitigate the effects of BPs during VQA training. In their method, the VQC consists of two components: a classical data encoding part and a variational part. A fixed classical input vector is fed into the encoding component, which uses the qubit encoding scheme to produce a quantum state that encodes the information. This quantum state is then evolved under the variational part, where the variational parameters are iteratively updated to minimize the cost function. The variational part is structured with alternating  layers of y-axis rotation gates and nearest-neighbour CNOT gates arranged in a line topology.

Through a series of numerical experiments, the authors examined the convergence speed of VQCs with and without NN-generated parameters across various numbers of qubits. The results demonstrate that VQCs using NN-derived parameters reach the target cost function value in significantly fewer iterations. This finding suggests the potential of neural networks to enhance the trainability of VQCs.

\section{Method}
\label{Sect:method}

In VQAs, the encoding step typically involves converting classical input data into a specific quantum state using fixed encoding rules, which are often tailored to the algorithm's requirements. In Ref. \cite{abp},
the classical input data is mapped into a quantum state by the following formula:
\begin{align}
|x_i\rangle = \bigotimes_j \cos(x_i^j)|0\rangle + \sin(x_i^j)|1\rangle,
\end{align}
where $x_i$ represents the $i$-th sample, while $x_i^j$ denotes the $j$-th feature of that sample. The input data is fixed at the value $x=(\frac{\pi}{4},\frac{\pi}{4},\ldots,\frac{\pi}{4})$, thus the input quantum state is the balanced superposition state $|+\rangle^{\otimes N}$.

To further explore the effectiveness of using neural networks to generate VQC's parameters, we remove the data encoding part, allowing the input of the variational part to be an arbitrary quantum state. These quantum states are sampled from the Haar measure and can be regarded as a form of quantum data, as well as quantum states that encode classical information. This modification broadens the applicability of the VQC, extending its utility beyond classical data processing to encompass both quantum and classical data processing.

\begin{figure}[!h]
    \centering
    \includegraphics[width=0.5\textwidth]{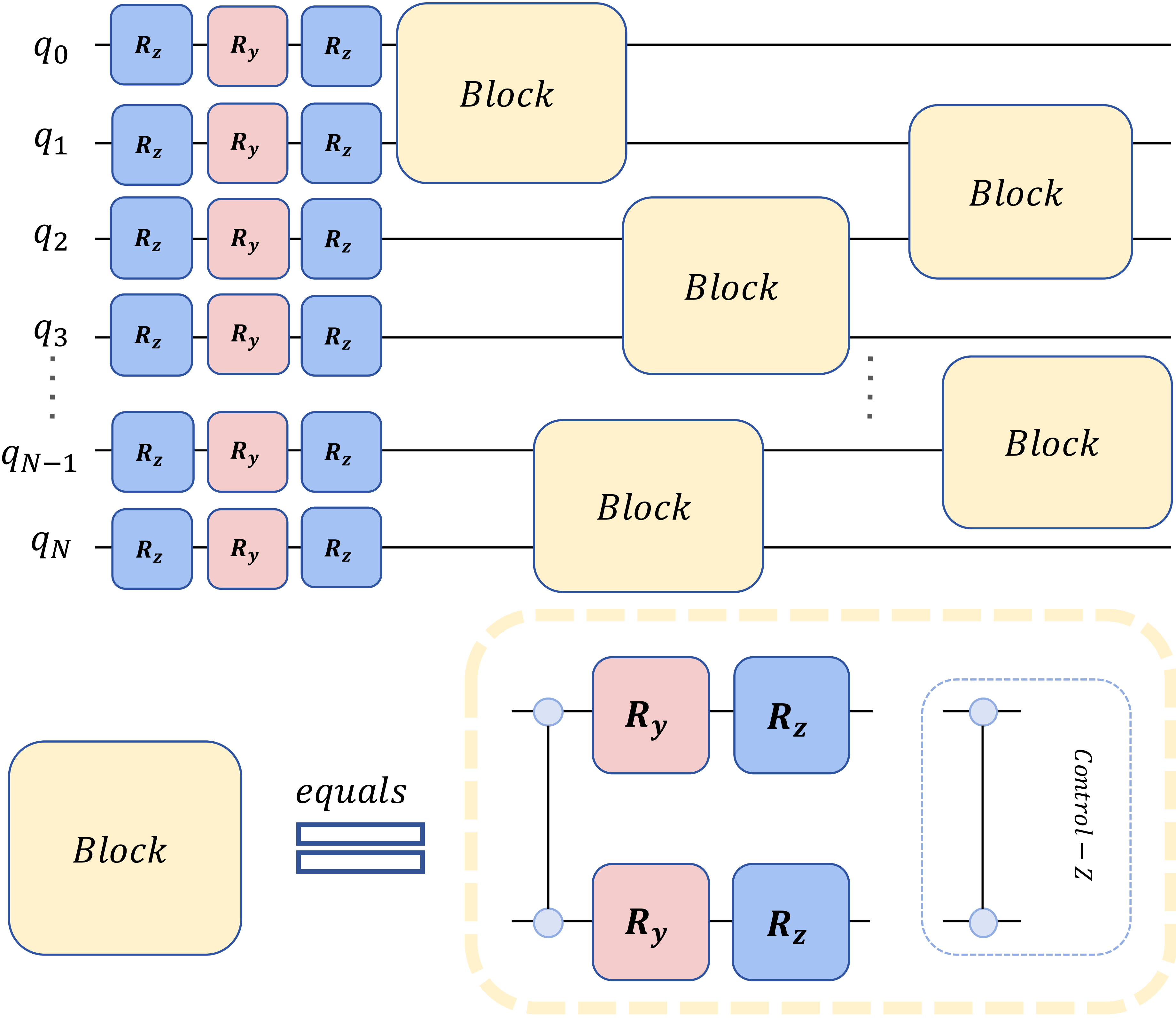}
    \caption{The construction of quantum circuits with random structures}
    \label{fig:circuit&block}
\end{figure}

The structure of the VQC is also expanded, transitioning from a fixed configuration to a randomized one. To construct a random circuit, we begin by applying a one-qubit unitary operation $U$ to each qubit. This unitary operation can be decomposed as $U=R_z(\theta_1)R_y(\theta_2)R_z(\theta_3)$ \cite{book}. Subsequently, we randomly select $L$ pairs of qubits and introduce a block of gates on each pair of qubits. Each block consists of a controlled-Z ($CZ$) gate connecting the selected qubits, followed by additional $R_y$ and $R_z$ gates on the same qubits. The structures of a VQC and a block are depicted in Figure \ref{fig:circuit&block}.

This method of adding blocks is equivalent to alternately adding one $CZ$ gate and one layer of $N$ one-qubit unitary operations on $N$ qubits. The equivalent VQC can be represented as:
\begin{align}
\mathcal{U}=\prod_{l=1}^L  \big(\bigotimes_{i=1}^N U^{(i)} CZ_l\big)   \bigotimes_{i=1}^N U^{(i)},
\end{align}
where $L$ denotes the number of blocks, $N$ represents the number of qubits, and the superscript $(i)$ refers to the index of the target qubit for a one-qubit unitary operation $U$. If $U=R_z(\theta_1)R_y(\theta_2)R_z(\theta_3)$ is placed after a $CZ$ gate, the first $R_z$ commutes with $CZ$ and can be moved before the $CZ$ gate, where it can be merged with the preceding one-qubit unitary operation. If $U$ is placed without a preceding $CZ$ gate, it can still be merged with the previous one-qubit unitary operation. Repeat the commutation and merge from the last block to the first block, we have
\begin{align}
\mathcal{U}=\prod_{l=1}^L  \big(\bigotimes_{i\in q(l)} R_z^{(i)}R_y^{(i)} CZ_l\big)   \bigotimes_{i=1}^N R_z^{(i)}R_y^{(i)}R_z^{(i)},
\end{align}
where $q(l)$ denotes the two target qubits of the $l-$th $CZ$ gate. The structure described in the above formula is precisely the random VQC structure we construct. Each rotation gate $R_z (R_y)$ has its own angle as a parameter, so the total number of parameters in the VQC is $3N+4L$.

\begin{figure}[!h]
\centering
\subfigure[SQC]{
\label{Fig.SQC}
\includegraphics[width=\textwidth]{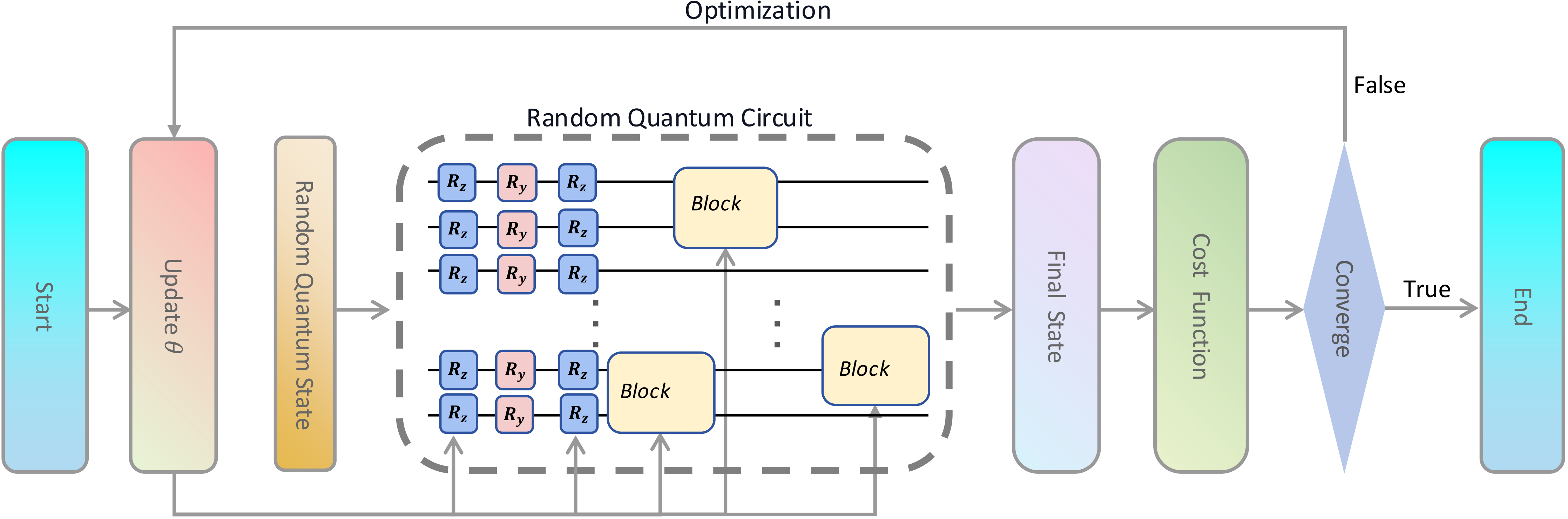}}
\subfigure[NEQC]{
\label{Fig.NEQC}
\includegraphics[width=\textwidth]{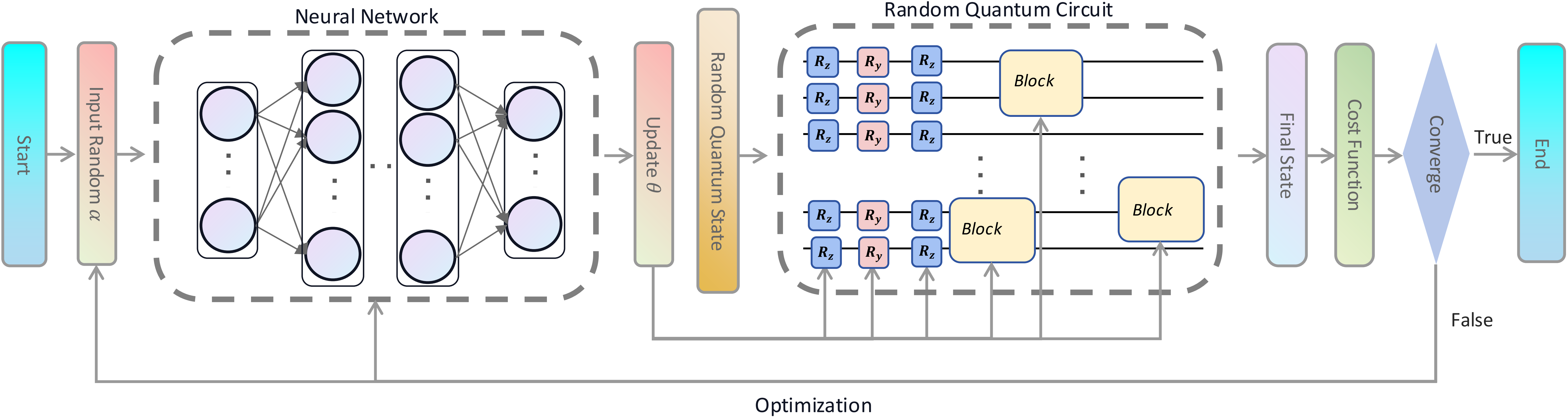}}
\caption{Illustration of the SQC and NEQC models. (a) The circuit parameters in the Standard Quantum Circuit (SQC) are directly initialized and iteratively updated. (b) The circuit parameters in the Neural Enhanced Quantum Circuit (NEQC) are derived from the neural network. The input vector and the parameters of the neural network are iteratively updated.  }
\label{corqc}
\end{figure}

The training process is illustrated in Figure \ref{corqc}. The SQC and NEQC models stand for the Standard Quantum Circuit and the Neural Enhanced Quantum Circuit, respectively.
For the SQC model, the quantum circuit parameters, denoted as $\theta$, are initialized set to random values and then iteratively updated. If the circuit consists of $N$ qubits and $L$ blocks, the dimension of $\theta$ is $3N+4L$. The gradients of $\theta$ are calculated based on the output quantum state and the cost function. These gradients are then used to update $\theta$ via a gradient-based optimizer. This process is repeated until the cost function converges or a maximum number of iterations is reached.
In contrast, for the NEQC model, the process begins with the random generation of an input vector, denoted as $\alpha$, which serves as the input for the neural network. The output of the neural network, $\theta$, is used to parameterize the random quantum circuit. The gradients of $\alpha$ and the neural network parameters $w$ are then calculated to update both $\alpha$ and $w$.

The architecture of the neural network is detailed in Table \ref{architectures}. The architecture in the left column closely resembles one of the three neural network architectures used in the previous work \cite{abp}. Since those three architectures demonstrate comparable performance, we opt to use only one of them in this study, which we refer to as NEQC-NN. We propose a network architecture, named NEQC-CNN, which is described in the right column. This architecture employs 1-dimensional convolutional layer, designed for sequential data such as time series and signals. The first and second arguments of the convolution operation represent the number of input and output channels, respectively. The kernel size is set to 3, the stride to 1, and the padding to 1. The input size of the network is 4, consistent with the input size of the NEQC-NN model. The first $3N+4L$ elements of the output are used to parameterize the quantum circuit.

\begin{table}[!h]
\centering
\caption{The architecture of the neural network}
\begin{tabular}{cc}
\toprule
NEQC-NN & NEQC-CNN\\
\midrule
Linear(4,10) & Conv1d(1,10)\\
Tanh & Tanh\\
Linear(10,20) & Conv1d(10,20)\\
Tanh & Tanh\\
Linear(20,$3N+4L$) & Conv1d(20,$\lfloor\frac{3N+4L}{4}\rfloor+1$)\\
Tanh & Tanh\\
\bottomrule
\end{tabular}
\label{architectures}
\end{table}

The cost function used in this study is identical to that of the previous work \cite{abp}. It is defined by
\begin{align}\label{Eq: cost function}
    C = 1 - \frac{1}{N}\sum_i Tr[(\ket{0}\bra{0}_i\otimes I_{\overline{i}})\rho ],
\end{align}
where $\ket{0}\bra{0}_i$ acts on the $i$-th qubit, $I_{\overline{i}}$ acts on all qubits except the $i$-th, and $\rho$ is the output state of the quantum circuit. The cost function $C=0$ if and only if all single-qubit measurements yield the the outcome 0. Minimizing this cost function is equivalent to training the quantum circuit to transform an arbitrary input quantum state $|\phi\rangle$ into $|0\rangle^{\otimes N}$. The inverse of this circuit can then convert $|0\rangle^{\otimes N}$ into the target quantum state $|\phi\rangle$, effectively learning a quantum state preparation task. It is important to note that in both QAOA and VQE, the objective is similarly to train the VQC to learn a specific quantum state (the ground state of a Hamiltonian).

To evaluate the effectiveness of the NN-based BP mitigation approach, we compare the SQC, NEQC-NN and NEQC-CNN models based on the number of iterations required for the cost function to converge. Additionally, we visualize the loss landscape to observe its smoothness. We also compare the expressibility, which is closely related to the trainability, between the three models.

We use the visualization technique proposed in Ref. \cite{losslandscape} to plot the loss landscape. First, we  sample 200 points within the interval $[-0.5, 0.5]$ on both the x-axis and y-axis to form a grid. We correspond the center point of the grid $(x=0,y=0)$ to the optimized parameters of the model. For the SQC model, the optimized parameters are the rotation angles $\theta$ within the VQC. For the NEQC-NN and NEQC-CNN models, the optimized parameters include the input vector $\alpha$ for the neural network as well as the neural network's adjustable parameters $w$. The optimized parameters are flattened into a vector, denoted as $Params(0,0)$, and normalized.

To explore the high-dimensional space of parameters, we need to generate two random direction vectors, denoted as $d_1$ and $d_2$, which are also normalized and share the same dimension as $Params(0,0)$.
The model parameters corresponding to each grid point $(x,y)$ are then calculated as:
\begin{align}\label{Eq: newparams}
    Params(x,y) = Params(0,0) + x \cdot d_1 + y \cdot d_2.
\end{align}
The resulting parameter vector $Params(x,y)$  for each grid point is reshaped to match the model's structure, allowing the corresponding loss to be computed using these parameters. The loss value serves as the $z$-coordinate, forming a 3D plot together with the $(x, y)$ coordinates.

We analyze the variation in expressibility when transitioning from the SQC model to the NEQC-NN and NEQC-CNN models. The estimation of expressibility for the SQC model follows the three steps procedure outlined in Section \ref{sect:express}. For the NEQC-NN and NEQC-CNN models, the process is similar, with a slight modification in the first step. Specifically, $2K$ random vectors are uniformly sampled from the interval $[0,2\pi]$. Each vector is used as input to the neural network with optimized $w$ and the resulting output vector is employed to parameterize the VQC, generating a corresponding quantum state. In this manner, $2K$ quantum states are produced by initializing the VQC with $\ket{0}$ as the input and using $2K$ sets of parameters derived from the neural network.

\section{Numerical results}
\label{Sect:experiment}

We first describe the settings of our numerical experiment. The qubit count $N$ of the VQCs varies from 3 to 8, while the number of blocks $L$ is defined as $L=\lceil N^2\ln N\rceil$. For each qubit count, the SQC, NEQC-NN and NEQC-CNN models are trained for 10 independent runs, each initialized with different parameters. In every run, the structure of the VQC and the input quantum state are randomly generated. We employ stochastic gradient descent (SGD) with momentum as the optimizer, which works well for our models. The training is terminated when the loss value falls below 0.001. Every 100 iterations, the change in loss is calculated, and if it is smaller than 0.0001, the training is halted. The models are constructed and trained using the PyTorch \cite{pytorch} and TensorCircuit \cite{tensorcircuit} frameworks.

\begin{figure}[hbtp]
\centering
\subfigure[3 qubits]
{
    \begin{minipage}[b]{.3\linewidth}
        \centering
        \includegraphics[scale=0.385]{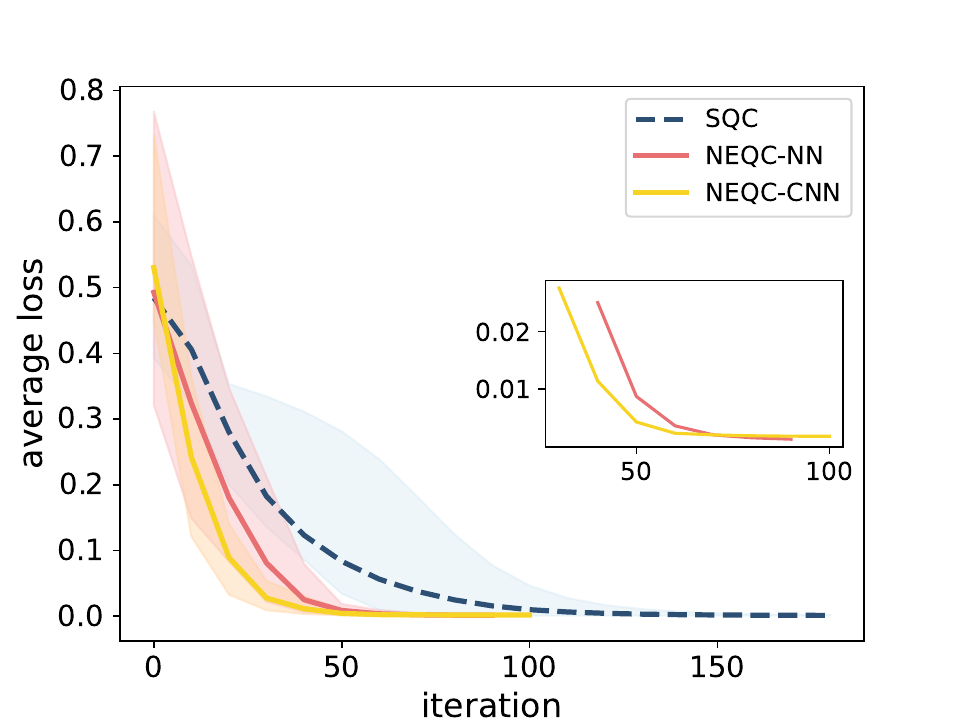}
    \end{minipage}
}
\subfigure[4 qubits]
{
    \begin{minipage}[b]{.3\linewidth}
        \centering
        \includegraphics[scale=0.385]{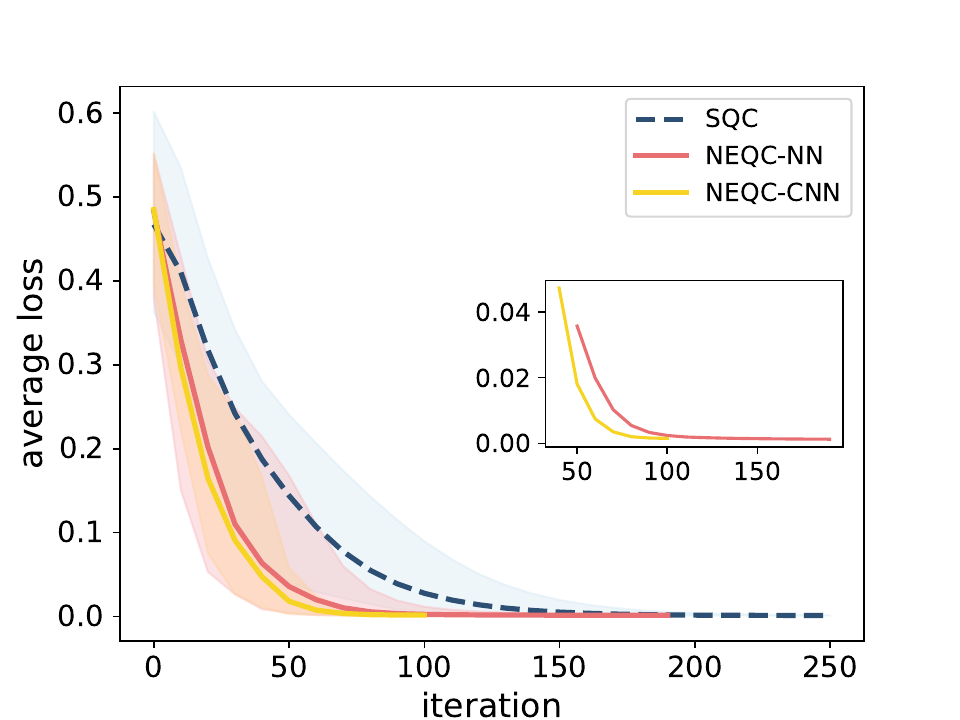}
    \end{minipage}
}
\subfigure[5 qubits]
{
    \begin{minipage}[b]{.3\linewidth}
        \centering
        \includegraphics[scale=0.385]{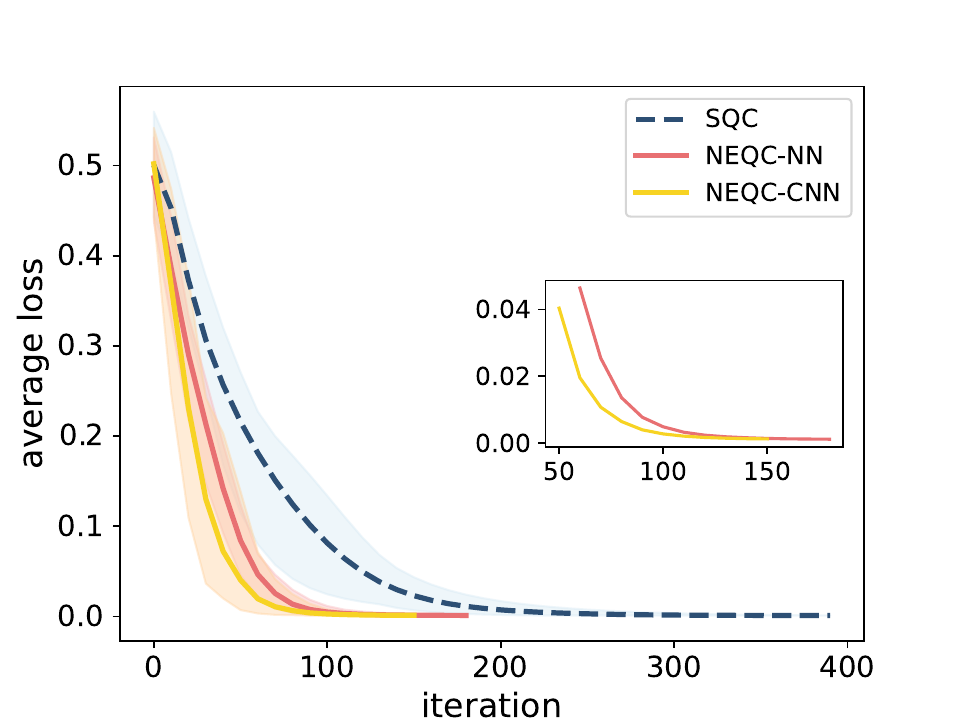}
    \end{minipage}
}
\subfigure[6 qubits]
{
    \begin{minipage}[b]{.3\linewidth}
        \centering
        \includegraphics[scale=0.385]{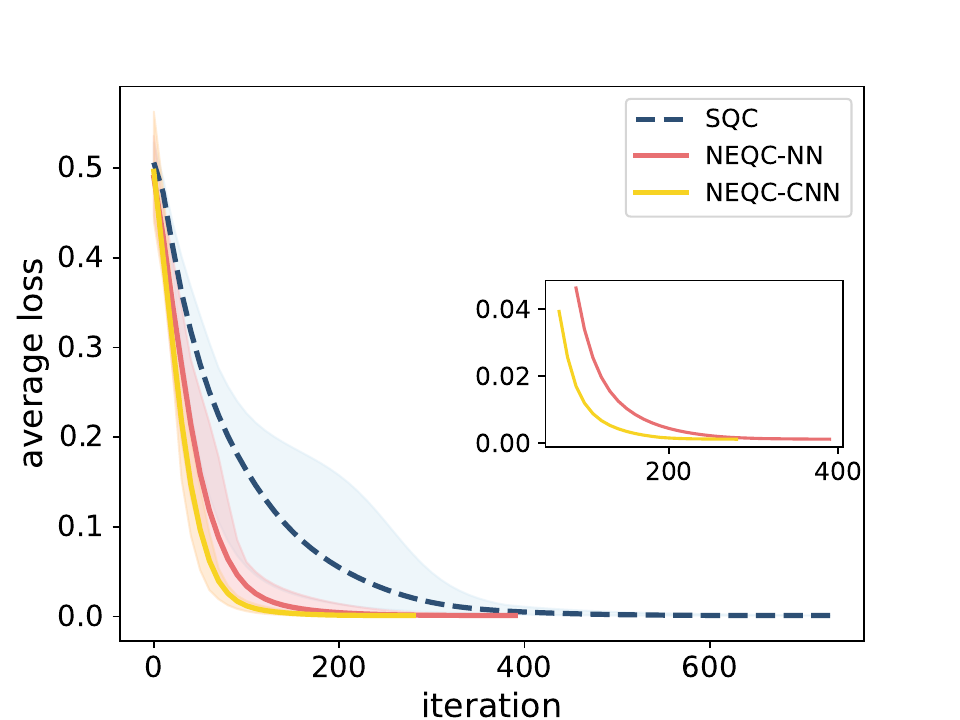}
    \end{minipage}
}
\subfigure[7 qubits]
{
    \begin{minipage}[b]{.3\linewidth}
        \centering
        \includegraphics[scale=0.385]{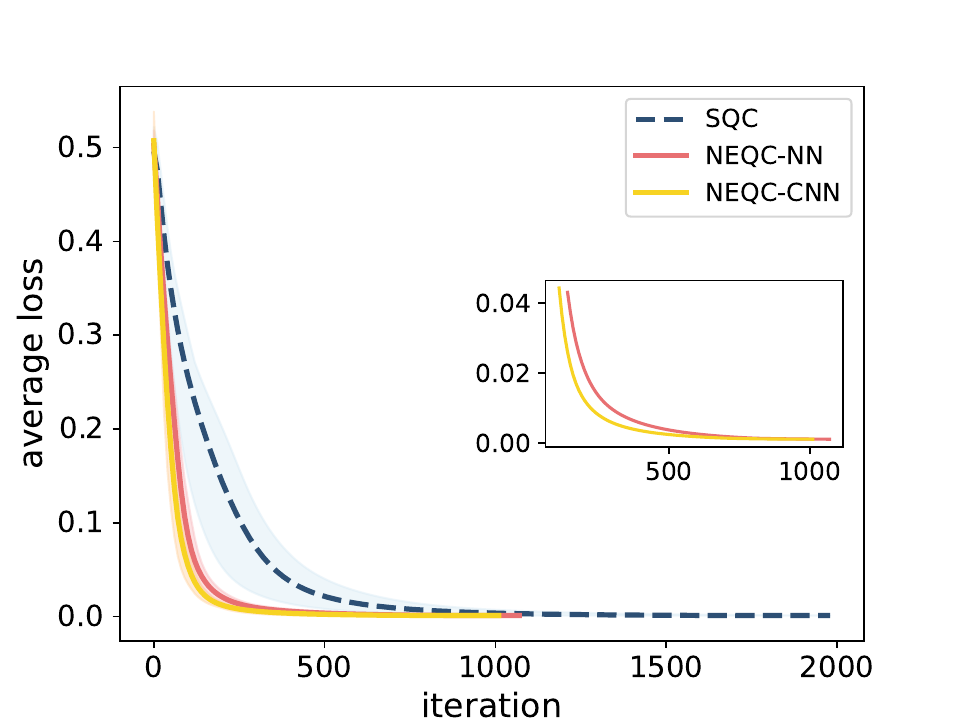}
    \end{minipage}
}
\subfigure[8 qubits]
{
    \begin{minipage}[b]{.3\linewidth}
        \centering
        \includegraphics[scale=0.385]{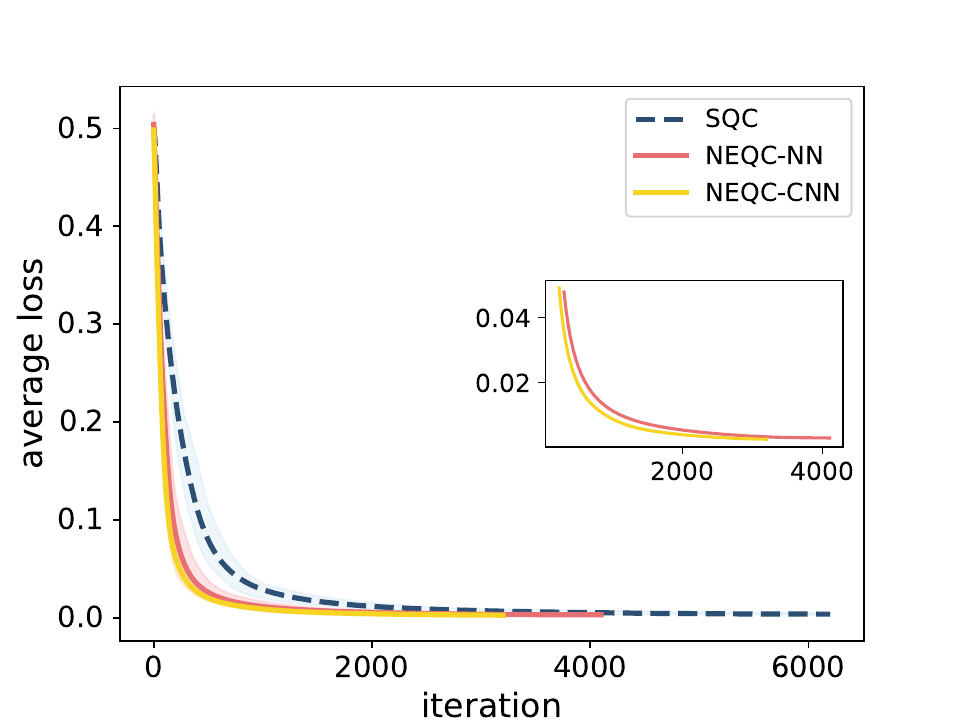}
    \end{minipage}
}
\caption{The variation of average loss with respect to the number of iterations. }
\label{loss_vs_iteration}
\end{figure}

Figure \ref{loss_vs_iteration} illustrates the variation in average loss as a function of the number of iterations. The blue, red and yellow lines corresponds to the SQC, NEQC-NN and NEQC-CNN models, respectively.  The average loss is computed over 10 independent runs and plotted every 10 iterations. The shaded regions represent the range between the maximum and minimum loss values. The results demonstrate that, for the same number of qubits, the NEQC models reach the global minimum in fewer iterations compared to the SQC model. While the number of iterations needed to achieve the global minimum increases with the number of qubits, this increase is less pronounced for the NEQC models. These findings suggest that incorporating a neural network for parameter generation effectively mitigates the BP problem when training VQCs with random structures and inputs.

Table \ref{convergence_iteration} illustrates the number of iterations required for training to converge across different qubit counts and model configurations. The number of iterations for the NEQC-NN model to converge is 45\%-67\% of that of the SQC model, while the number of iterations for the NEQC-CNN model to converge is further reduced, reaching only 36\%-58\% of the SQC model. When the number of qubits is 8, all runs terminate due to reaching the convergence condition, rather than achieving a loss value below 0.001. Table \ref{convergence_loss} presents the loss values at the point of convergence for different qubit counts and model configurations. We can see that all models converge to comparable loss values. For 3 to 7 qubits, the models achieve an optimal loss value, while for 8 qubits, the training converges to a loss value slightly larger than 0.001.

\begin{table}[!htbp]
    \centering
    \resizebox{0.9999\textwidth}{!}{
    \begin{threeparttable}
    \caption{Number of iterations at convergence for varying qubit counts and model configurations}

    \begin{tabular}{lllllllllllll}
    \toprule
         qubit count& model & run 1 & run 2 & run 3 & run 4 & run 5 & run 6 & run 7 & run 8 & run 9 & run 10 & average \\
        \midrule
        \multirow{3}{*}{3}  &  \multirow{1}{*}{SQC} &
        184 &
        109 &
        135 &
        118 &
        106 &
        90  &
        143 &
        171 &
        183 &
        143 &
        138
        \\
        ~ & \multirow{1}{*}{NEQC-NN}&
        \multirow{1}{*}{76} &
        \multirow{1}{*}{64} &
        \multirow{1}{*}{90} &
        \multirow{1}{*}{71} &
        \multirow{1}{*}{72} &
        \multirow{1}{*}{64} &
        \multirow{1}{*}{93} &
        \multirow{1}{*}{69} &
        \multirow{1}{*}{67} &
        \multirow{1}{*}{54} &
        \multirow{1}{*}{72}
        \\
        ~ & \multirow{1}{*}{NEQC-CNN}&
        \multirow{1}{*}{61} &
        \multirow{1}{*}{64} &
        \multirow{1}{*}{104}&
        \multirow{1}{*}{62} &
        \multirow{1}{*}{58} &
        \multirow{1}{*}{68} &
        \multirow{1}{*}{67} &
        \multirow{1}{*}{55} &
        \multirow{1}{*}{56} &
        \multirow{1}{*}{96} &
        \multirow{1}{*}{69}
        \\
        \\

        \multirow{3}{*}{4}  & \multirow{1}{*}{SQC} &
        193 &
        255 &
        186 &
        187 &
        145 &
        182 &
        211 &
        216 &
        137 &
        154 &
        187
        \\
        ~ & \multirow{1}{*}{NEQC-NN}&
        \multirow{1}{*}{86} &
        \multirow{1}{*}{99} &
        \multirow{1}{*}{82} &
        \multirow{1}{*}{91} &
        \multirow{1}{*}{73} &
        \multirow{1}{*}{71} &
        \multirow{1}{*}{80} &
        \multirow{1}{*}{69} &
        \multirow{1}{*}{194} &
        \multirow{1}{*}{114} &
        \multirow{1}{*}{96}
        \\
        ~ & \multirow{1}{*}{NEQC-CNN}&
        \multirow{1}{*}{81} &
        \multirow{1}{*}{78} &
        \multirow{1}{*}{84} &
        \multirow{1}{*}{67} &
        \multirow{1}{*}{66} &
        \multirow{1}{*}{86} &
        \multirow{1}{*}{59} &
        \multirow{1}{*}{90} &
        \multirow{1}{*}{108} &
        \multirow{1}{*}{71} &
        \multirow{1}{*}{79}
        \\
        \\

        \multirow{3}{*}{5}  & \multirow{1}{*}{SQC} &
        299 &
        249 &
        391 &
        225 &
        319 &
        323 &
        326 &
        282 &
        368 &
        283 &
        307
        \\
        ~ & \multirow{1}{*}{NEQC-NN}&
        \multirow{1}{*}{97} &
        \multirow{1}{*}{121} &
        \multirow{1}{*}{127} &
        \multirow{1}{*}{181} &
        \multirow{1}{*}{182} &
        \multirow{1}{*}{155} &
        \multirow{1}{*}{130} &
        \multirow{1}{*}{111} &
        \multirow{1}{*}{169} &
        \multirow{1}{*}{116} &
        \multirow{1}{*}{139}
        \\
        ~ & \multirow{1}{*}{NEQC-CNN}&
        \multirow{1}{*}{79} &
        \multirow{1}{*}{125} &
        \multirow{1}{*}{152} &
        \multirow{1}{*}{112} &
        \multirow{1}{*}{90} &
        \multirow{1}{*}{105} &
        \multirow{1}{*}{99} &
        \multirow{1}{*}{149} &
        \multirow{1}{*}{93} &
        \multirow{1}{*}{90} &
        \multirow{1}{*}{109}
        \\
        \\

        \multirow{3}{*}{6}  & \multirow{1}{*}{SQC} &
        533 &
        634 &
        592 &
        735 &
        552 &
        507 &
        497 &
        608 &
        572 &
        591 &
        582
        \\
        ~ & \multirow{1}{*}{NEQC-NN}&
        \multirow{1}{*}{314} &
        \multirow{1}{*}{393} &
        \multirow{1}{*}{266} &
        \multirow{1}{*}{255} &
        \multirow{1}{*}{219} &
        \multirow{1}{*}{311} &
        \multirow{1}{*}{274} &
        \multirow{1}{*}{179} &
        \multirow{1}{*}{194} &
        \multirow{1}{*}{285} &
        \multirow{1}{*}{269}
        \\
        ~ & \multirow{1}{*}{NEQC-CNN}&
        \multirow{1}{*}{225} &
        \multirow{1}{*}{197} &
        \multirow{1}{*}{197} &
        \multirow{1}{*}{207} &
        \multirow{1}{*}{213} &
        \multirow{1}{*}{160} &
        \multirow{1}{*}{199} &
        \multirow{1}{*}{196} &
        \multirow{1}{*}{231} &
        \multirow{1}{*}{285} &
        \multirow{1}{*}{211}
        \\
        \\

       \multirow{3}{*}{7}  & \multirow{1}{*}{SQC} &
        1707 &
        1402 &
        1986 &
        1146 &
        1571 &
        1128 &
        1735 &
        1338 &
        1621 &
        1713 &
        1535
        \\
       ~ & \multirow{1}{*}{NEQC-NN}&
        \multirow{1}{*}{1075} &
        \multirow{1}{*}{798} &
        \multirow{1}{*}{840} &
        \multirow{1}{*}{800} &
        \multirow{1}{*}{819} &
        \multirow{1}{*}{892} &
        \multirow{1}{*}{747} &
        \multirow{1}{*}{974} &
        \multirow{1}{*}{810} &
        \multirow{1}{*}{1066} &
        \multirow{1}{*}{882}
        \\
        ~ & \multirow{1}{*}{NEQC-CNN}&
        \multirow{1}{*}{841} &
        \multirow{1}{*}{595} &
        \multirow{1}{*}{711} &
        \multirow{1}{*}{790} &
        \multirow{1}{*}{899} &
        \multirow{1}{*}{459} &
        \multirow{1}{*}{672} &
        \multirow{1}{*}{735} &
        \multirow{1}{*}{1012} &
        \multirow{1}{*}{706} &
        \multirow{1}{*}{742}
        \\
        \\

        \multirow{3}{*}{8}  & \multirow{1}{*}{SQC} &
        6200 &
        5500 &
        5600 &
        5200 &
        4400 &
        4800 &
        5700 &
        4600 &
        3800 &
        5400 &
        5120
        \\
        ~ & \multirow{1}{*}{NEQC-NN}&
        \multirow{1}{*}{3800} &
        \multirow{1}{*}{3900} &
        \multirow{1}{*}{3400} &
        \multirow{1}{*}{3500} &
        \multirow{1}{*}{4100} &
        \multirow{1}{*}{3300} &
        \multirow{1}{*}{3400} &
        \multirow{1}{*}{3100} &
        \multirow{1}{*}{2800} &
        \multirow{1}{*}{3100} &
        \multirow{1}{*}{3440}
        \\
        ~ & \multirow{1}{*}{NEQC-CNN}&
        \multirow{1}{*}{3100} &
        \multirow{1}{*}{3100} &
        \multirow{1}{*}{2700} &
        \multirow{1}{*}{2400} &
        \multirow{1}{*}{3000} &
        \multirow{1}{*}{3200} &
        \multirow{1}{*}{3100} &
        \multirow{1}{*}{2800} &
        \multirow{1}{*}{2900} &
        \multirow{1}{*}{3200} &
        \multirow{1}{*}{2950}
        \\
        \bottomrule
    \end{tabular}
\label{convergence_iteration}
\end{threeparttable}
}
\end{table}

\begin{table}[!htbp]
    \centering
    \resizebox{0.9999\textwidth}{!}{
    \begin{threeparttable}
    \caption{Loss values at convergence for varying qubit counts and model configurations}

    \begin{tabular}{lllllllllllll}
    \toprule
         qubit count& model & run 1 & run 2 & run 3 & run 4 & run 5 & run 6 & run 7 & run 8 & run 9 & run 10 & average \\
        \midrule
        \multirow{3}{*}{3}  & \multirow{1}{*}{SQC} &
        $9.93\times10^{-4}$ &
        $9.48\times10^{-4}$ &
        $9.74\times10^{-4}$ &
        $9.94\times10^{-4}$ &
        $9.98\times10^{-4}$ &
        $9.80\times10^{-4}$&
        $9.69\times10^{-4}$ &
        $9.80\times10^{-4}$ &
        $9.67\times10^{-4}$ &
        $9.99\times10^{-4}$ &
        $9.80\times10^{-4}$
        \\
        ~ & \multirow{1}{*}{NEQC-NN}&
        \multirow{1}{*}{$9.39\times10^{-4}$} &
        \multirow{1}{*}{$9.30\times10^{-4}$} &
        \multirow{1}{*}{$9.35\times10^{-4}$} &
        \multirow{1}{*}{$8.94\times10^{-4}$} &
        \multirow{1}{*}{$9.37\times10^{-4}$} &
        \multirow{1}{*}{$8.72\times10^{-4}$} &
        \multirow{1}{*}{$9.83\times10^{-4}$} &
        \multirow{1}{*}{$8.93\times10^{-4}$} &
        \multirow{1}{*}{$9.07\times10^{-4}$} &
        \multirow{1}{*}{$9.61\times10^{-4}$} &
        \multirow{1}{*}{$9.25\times10^{-4}$}
        \\
        ~ & \multirow{1}{*}{NEQC-CNN}&
        \multirow{1}{*}{$9.55\times10^{-4}$} &
        \multirow{1}{*}{$9.04\times10^{-4}$} &
        \multirow{1}{*}{$9.96\times10^{-4}$}&
        \multirow{1}{*}{$8.53\times10^{-4}$} &
        \multirow{1}{*}{$8.38\times10^{-4}$} &
        \multirow{1}{*}{$8.89\times10^{-4}$} &
        \multirow{1}{*}{$8.62\times10^{-4}$} &
        \multirow{1}{*}{$9.46\times10^{-4}$} &
        \multirow{1}{*}{$9.37\times10^{-4}$} &
        \multirow{1}{*}{$9.78\times10^{-4}$} &
        \multirow{1}{*}{$9.16\times10^{-4}$}
        \\
        \\

        \multirow{3}{*}{4}  & \multirow{1}{*}{SQC} &
        $9.98\times10^{-4}$ &
        $9.92\times10^{-4}$ &
        $9.80\times10^{-4}$ &
        $9.80\times10^{-4}$ &
        $9.91\times10^{-4}$ &
        $9.95\times10^{-4}$&
        $9.74\times10^{-4}$ &
        $9.99\times10^{-4}$ &
        $9.93\times10^{-4}$ &
        $9.59\times10^{-4}$ &
        $9.86\times10^{-4}$
        \\
        ~ & \multirow{1}{*}{NEQC-NN}&
        \multirow{1}{*}{$9.65\times10^{-4}$} &
        \multirow{1}{*}{$9.42\times10^{-4}$} &
        \multirow{1}{*}{$9.67\times10^{-4}$} &
        \multirow{1}{*}{$9.83\times10^{-4}$} &
        \multirow{1}{*}{$9.61\times10^{-4}$} &
        \multirow{1}{*}{$9.92\times10^{-4}$} &
        \multirow{1}{*}{$9.48\times10^{-4}$} &
        \multirow{1}{*}{$9.59\times10^{-4}$} &
        \multirow{1}{*}{$9.86\times10^{-4}$} &
        \multirow{1}{*}{$9.46\times10^{-4}$} &
        \multirow{1}{*}{$9.65\times10^{-4}$}
        \\
        ~ & \multirow{1}{*}{NEQC-CNN}&
        \multirow{1}{*}{$9.61\times10^{-4}$} &
        \multirow{1}{*}{$9.56\times10^{-4}$} &
        \multirow{1}{*}{$9.54\times10^{-4}$} &
        \multirow{1}{*}{$8.85\times10^{-4}$} &
        \multirow{1}{*}{$8.36\times10^{-4}$} &
        \multirow{1}{*}{$9.00\times10^{-4}$} &
        \multirow{1}{*}{$9.78\times10^{-4}$} &
        \multirow{1}{*}{$9.55\times10^{-4}$} &
        \multirow{1}{*}{$9.83\times10^{-4}$} &
        \multirow{1}{*}{$9.76\times10^{-4}$} &
        \multirow{1}{*}{$9.83\times10^{-4}$}
        \\
        \\

        \multirow{3}{*}{5}  & \multirow{1}{*}{SQC} &
        $9.94\times10^{-4}$ &
        $9.83\times10^{-4}$ &
        $1.00\times10^{-3}$ &
        $9.80\times10^{-4}$ &
        $9.99\times10^{-4}$ &
        $9.88\times10^{-4}$ &
        $9.89\times10^{-4}$ &
        $9.81\times10^{-4}$ &
        $9.97\times10^{-4}$ &
        $9.87\times10^{-4}$ &
        $9.90\times10^{-4}$
        \\
        ~& \multirow{1}{*}{NEQC-NN}&
        \multirow{1}{*}{$9.70\times10^{-4}$} &
        \multirow{1}{*}{$9.75\times10^{-4}$} &
        \multirow{1}{*}{$9.72\times10^{-4}$} &
        \multirow{1}{*}{$9.92\times10^{-4}$} &
        \multirow{1}{*}{$9.82\times10^{-4}$} &
        \multirow{1}{*}{$9.89\times10^{-4}$} &
        \multirow{1}{*}{$9.79\times10^{-4}$} &
        \multirow{1}{*}{$9.70\times10^{-4}$} &
        \multirow{1}{*}{$9.76\times10^{-4}$} &
        \multirow{1}{*}{$9.67\times10^{-4}$} &
        \multirow{1}{*}{$9.77\times10^{-4}$}
        \\
        ~ & \multirow{1}{*}{NEQC-CNN}&
        \multirow{1}{*}{$9.88\times10^{-4}$} &
        \multirow{1}{*}{$9.83\times10^{-4}$} &
        \multirow{1}{*}{$9.60\times10^{-4}$} &
        \multirow{1}{*}{$9.81\times10^{-4}$} &
        \multirow{1}{*}{$9.71\times10^{-4}$} &
        \multirow{1}{*}{$9.64\times10^{-4}$} &
        \multirow{1}{*}{$9.91\times10^{-4}$} &
        \multirow{1}{*}{$9.68\times10^{-4}$} &
        \multirow{1}{*}{$9.95\times10^{-4}$} &
        \multirow{1}{*}{$9.66\times10^{-4}$} &
        \multirow{1}{*}{$9.77\times10^{-4}$}
        \\
        \\

        \multirow{3}{*}{6}  & \multirow{1}{*}{SQC} &
        $9.94\times10^{-4}$ &
        $9.90\times10^{-4}$ &
        $9.96\times10^{-4}$ &
        $9.98\times10^{-4}$ &
        $9.97\times10^{-4}$ &
        $9.95\times10^{-4}$ &
        $9.97\times10^{-4}$ &
        $9.98\times10^{-4}$ &
        $9.98\times10^{-4}$ &
        $9.96\times10^{-4}$ &
        $9.96\times10^{-4}$
        \\
        ~ & \multirow{1}{*}{NEQC-NN}&
        \multirow{1}{*}{$9.96\times10^{-4}$} &
        \multirow{1}{*}{$9.93\times10^{-4}$} &
        \multirow{1}{*}{$9.97\times10^{-4}$} &
        \multirow{1}{*}{$9.93\times10^{-4}$} &
        \multirow{1}{*}{$9.96\times10^{-4}$} &
        \multirow{1}{*}{$1.00\times10^{-3}$} &
        \multirow{1}{*}{$9.91\times10^{-4}$} &
        \multirow{1}{*}{$9.91\times10^{-4}$} &
        \multirow{1}{*}{$9.88\times10^{-4}$} &
        \multirow{1}{*}{$9.98\times10^{-4}$} &
        \multirow{1}{*}{$9.94\times10^{-4}$}
        \\
        ~ & \multirow{1}{*}{NEQC-CNN}&
        \multirow{1}{*}{$9.98\times10^{-4}$} &
        \multirow{1}{*}{$9.97\times10^{-4}$} &
        \multirow{1}{*}{$9.84\times10^{-4}$} &
        \multirow{1}{*}{$9.91\times10^{-4}$} &
        \multirow{1}{*}{$9.86\times10^{-4}$} &
        \multirow{1}{*}{$9.81\times10^{-4}$} &
        \multirow{1}{*}{$9.98\times10^{-4}$} &
        \multirow{1}{*}{$9.89\times10^{-4}$} &
        \multirow{1}{*}{$9.80\times10^{-4}$} &
        \multirow{1}{*}{$9.96\times10^{-4}$} &
        \multirow{1}{*}{$9.90\times10^{-4}$}
        \\
        \\

       \multirow{3}{*}{7}  & \multirow{1}{*}{SQC} &
        $9.99\times10^{-4}$ &
        $9.98\times10^{-4}$ &
        $1.00\times10^{-3}$ &
        $9.97\times10^{-4}$ &
        $9.98\times10^{-4}$ &
        $1.00\times10^{-3}$ &
        $9.99\times10^{-4}$ &
        $9.99\times10^{-4}$ &
        $9.96\times10^{-4}$ &
        $9.98\times10^{-4}$ &
        $9.98\times10^{-4}$
        \\
       ~ & \multirow{1}{*}{NEQC-NN}&
        \multirow{1}{*}{$9.97\times10^{-4}$} &
        \multirow{1}{*}{$9.98\times10^{-4}$} &
        \multirow{1}{*}{$9.97\times10^{-4}$} &
        \multirow{1}{*}{$1.00\times10^{-3}$} &
        \multirow{1}{*}{$9.95\times10^{-4}$} &
        \multirow{1}{*}{$9.98\times10^{-4}$} &
        \multirow{1}{*}{$9.97\times10^{-4}$} &
        \multirow{1}{*}{$9.98\times10^{-4}$} &
        \multirow{1}{*}{$1.00\times10^{-3}$} &
        \multirow{1}{*}{$9.95\times10^{-4}$} &
        \multirow{1}{*}{$9.98\times10^{-4}$}
        \\
        ~ & \multirow{1}{*}{NEQC-CNN}&
        \multirow{1}{*}{$9.99\times10^{-4}$} &
        \multirow{1}{*}{$9.99\times10^{-4}$} &
        \multirow{1}{*}{$9.98\times10^{-4}$} &
        \multirow{1}{*}{$9.99\times10^{-4}$} &
        \multirow{1}{*}{$9.99\times10^{-4}$} &
        \multirow{1}{*}{$9.98\times10^{-4}$} &
        \multirow{1}{*}{$9.98\times10^{-4}$} &
        \multirow{1}{*}{$9.98\times10^{-4}$} &
        \multirow{1}{*}{$1.00\times10^{-3}$} &
        \multirow{1}{*}{$9.98\times10^{-4}$} &
        \multirow{1}{*}{$9.99\times10^{-4}$}
        \\
        \\

        \multirow{3}{*}{8}  & \multirow{1}{*}{SQC} &
        $2.54\times10^{-3}$ &
        $5.18\times10^{-3}$ &
        $4.00\times10^{-3}$ &
        $3.32\times10^{-3}$ &
        $3.41\times10^{-3}$ &
        $3.38\times10^{-3}$ &
        $3.55\times10^{-3}$ &
        $3.94\times10^{-3}$ &
        $3.29\times10^{-3}$ &
        $3.63\times10^{-3}$ &
        $3.62\times10^{-3}$
        \\
        ~ & \multirow{1}{*}{NEQC-NN}&
        \multirow{1}{*}{$3.35\times10^{-3}$} &
        \multirow{1}{*}{$3.55\times10^{-3}$} &
        \multirow{1}{*}{$2.89\times10^{-3}$} &
        \multirow{1}{*}{$2.67\times10^{-3}$} &
        \multirow{1}{*}{$3.16\times10^{-3}$} &
        \multirow{1}{*}{$2.11\times10^{-3}$} &
        \multirow{1}{*}{$3.04\times10^{-3}$} &
        \multirow{1}{*}{$2.23\times10^{-3}$} &
        \multirow{1}{*}{$3.57\times10^{-3}$} &
        \multirow{1}{*}{$2.71\times10^{-3}$} &
        \multirow{1}{*}{$2.93\times10^{-3}$}
        \\
        ~ & \multirow{1}{*}{NEQC-CNN}&
        \multirow{1}{*}{$3.20\times10^{-3}$} &
        \multirow{1}{*}{$2.69\times10^{-3}$} &
        \multirow{1}{*}{$3.12\times10^{-3}$} &
        \multirow{1}{*}{$2.14\times10^{-3}$} &
        \multirow{1}{*}{$2.34\times10^{-3}$} &
        \multirow{1}{*}{$2.30\times10^{-3}$} &
        \multirow{1}{*}{$2.45\times10^{-3}$} &
        \multirow{1}{*}{$2.56\times10^{-3}$} &
        \multirow{1}{*}{$2.23\times10^{-3}$} &
        \multirow{1}{*}{$2.47\times10^{-3}$} &
        \multirow{1}{*}{$2.55\times10^{-3}$}
        \\
        \bottomrule
    \end{tabular}
\label{convergence_loss}
\end{threeparttable}
}
\end{table}

\begin{figure}[htbp]
\centering
\subfigure{\includegraphics[width=0.23\textwidth]{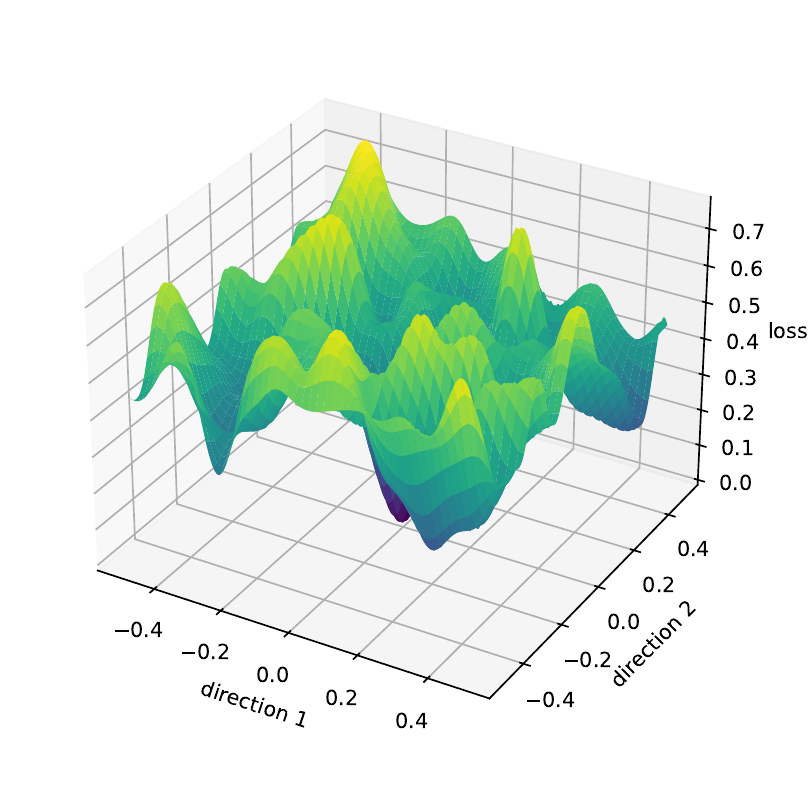}}
\subfigure{\includegraphics[width=0.23\textwidth]{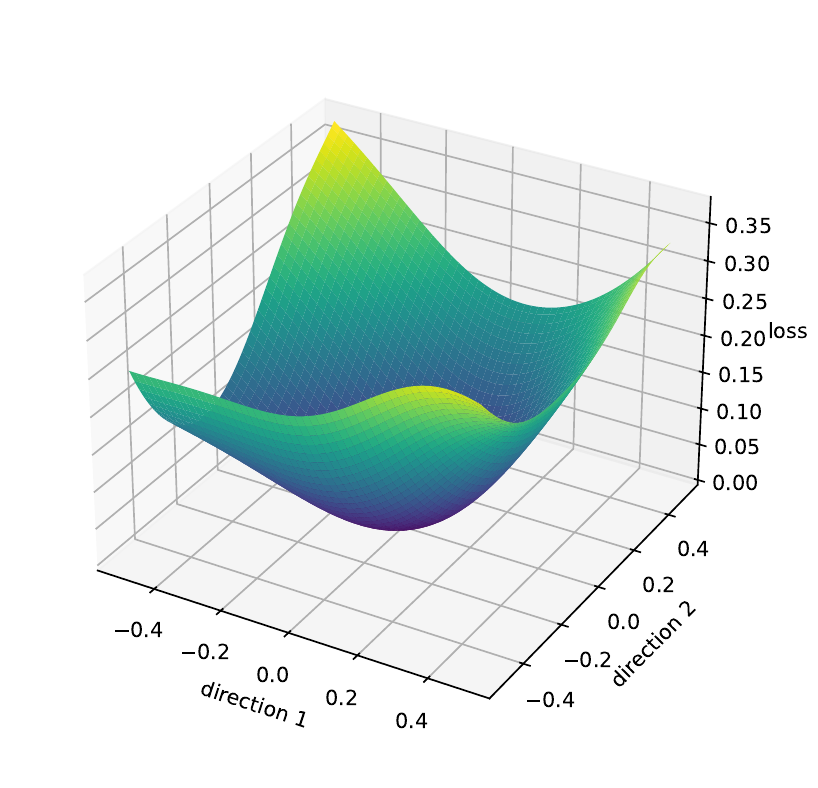}}
\subfigure{\includegraphics[width=0.23\textwidth]{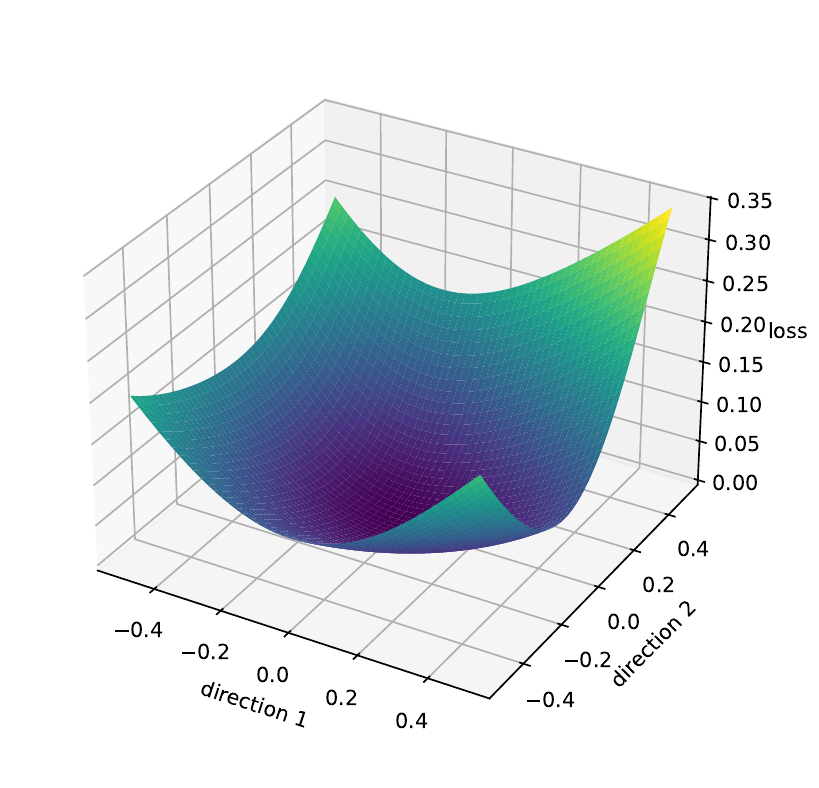}}\\
\vspace{-6mm}
\subfigure{\includegraphics[width=0.23\textwidth]{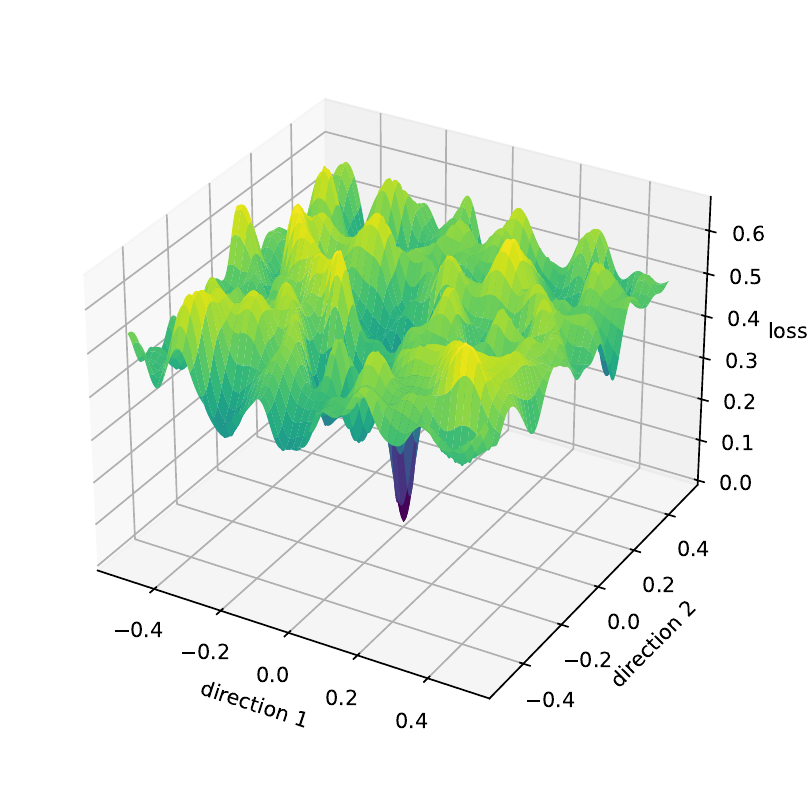}}
\subfigure{\includegraphics[width=0.23\textwidth]{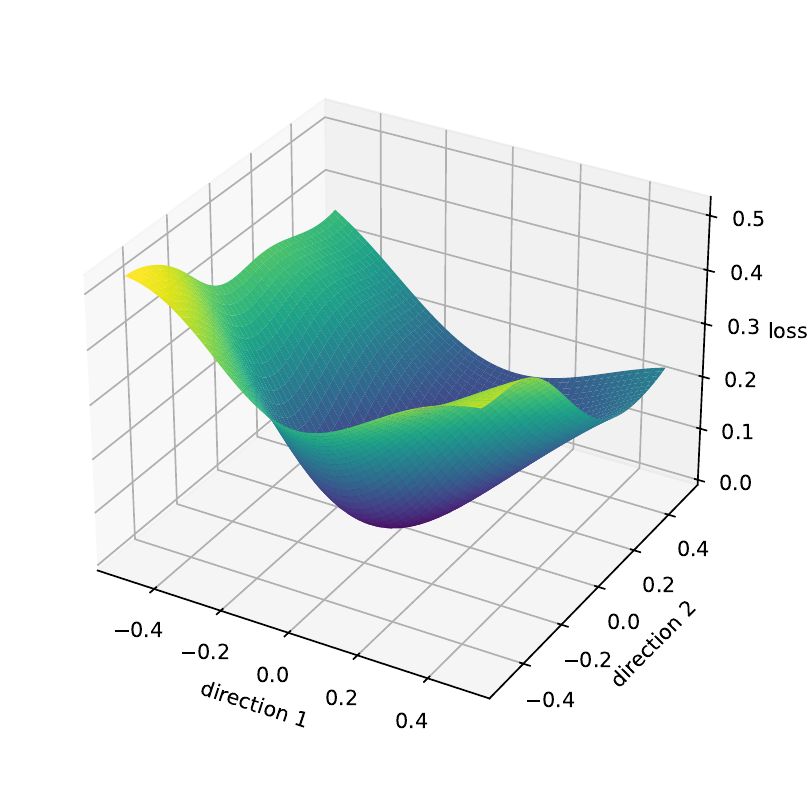}}
\subfigure{\includegraphics[width=0.23\textwidth]{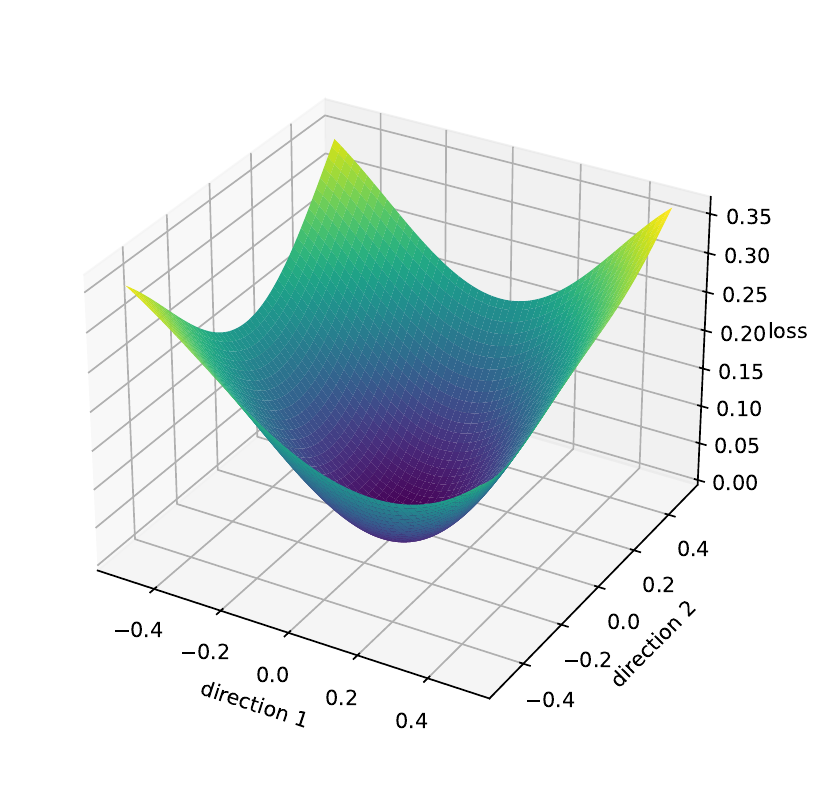}}\\
\vspace{-6mm}
\subfigure{\includegraphics[width=0.23\textwidth]{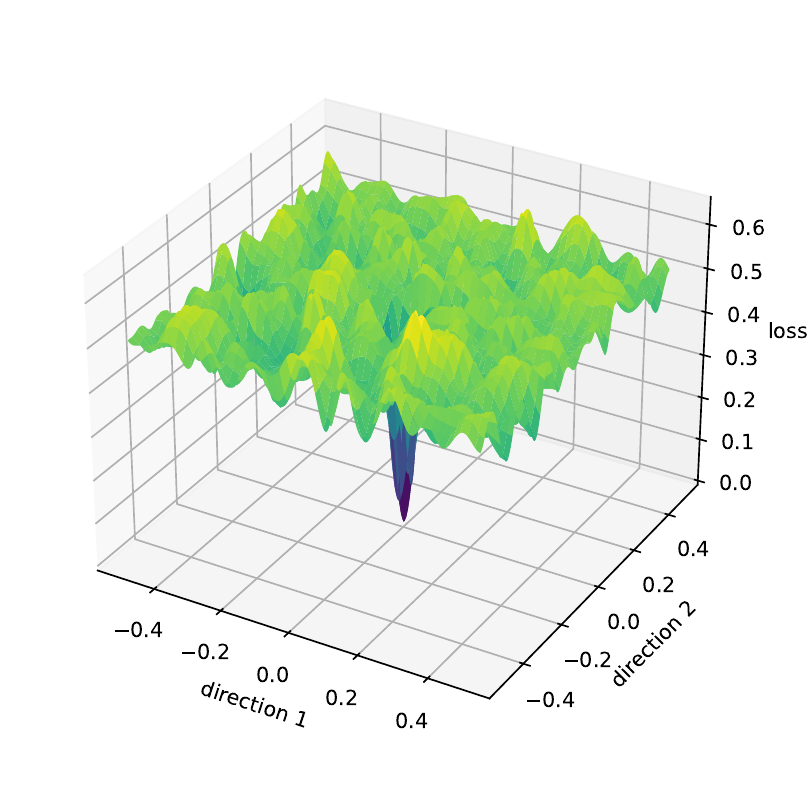}}
\subfigure{\includegraphics[width=0.23\textwidth]{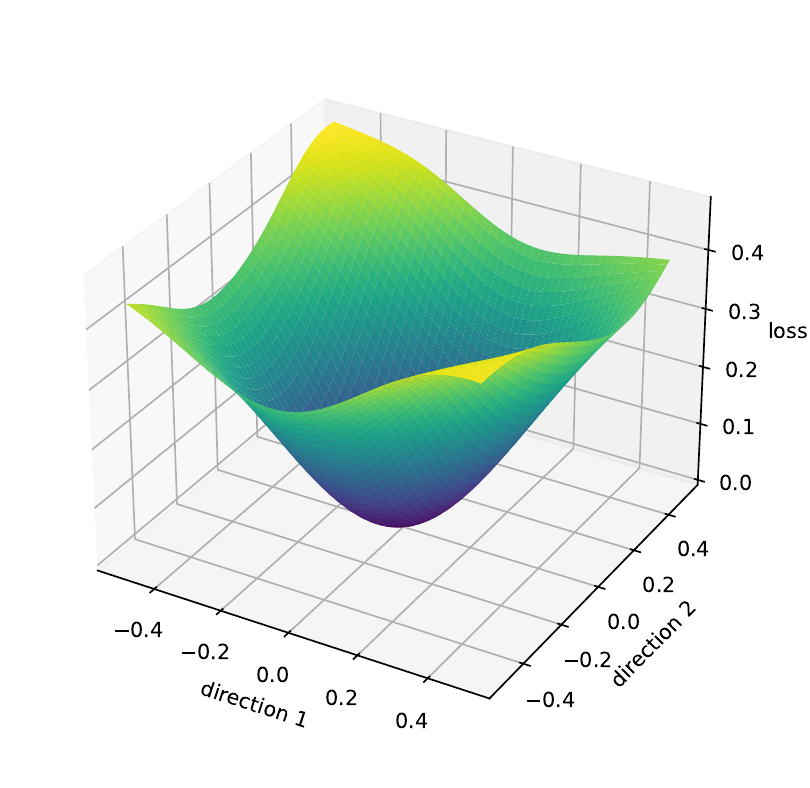}}
\subfigure{\includegraphics[width=0.23\textwidth]{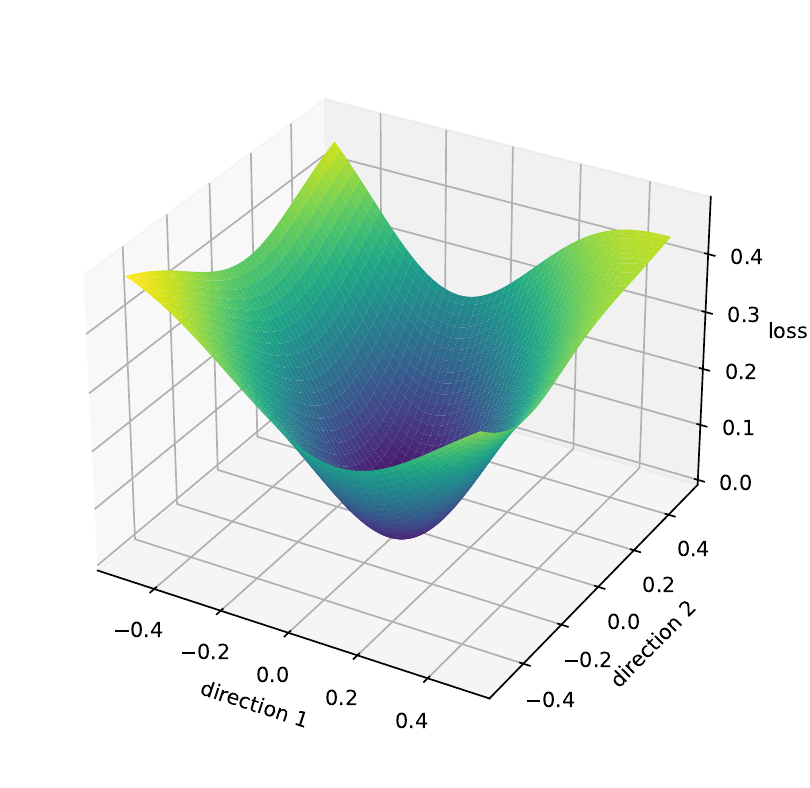}}\\
\vspace{-6mm}
\subfigure{\includegraphics[width=0.23\textwidth]{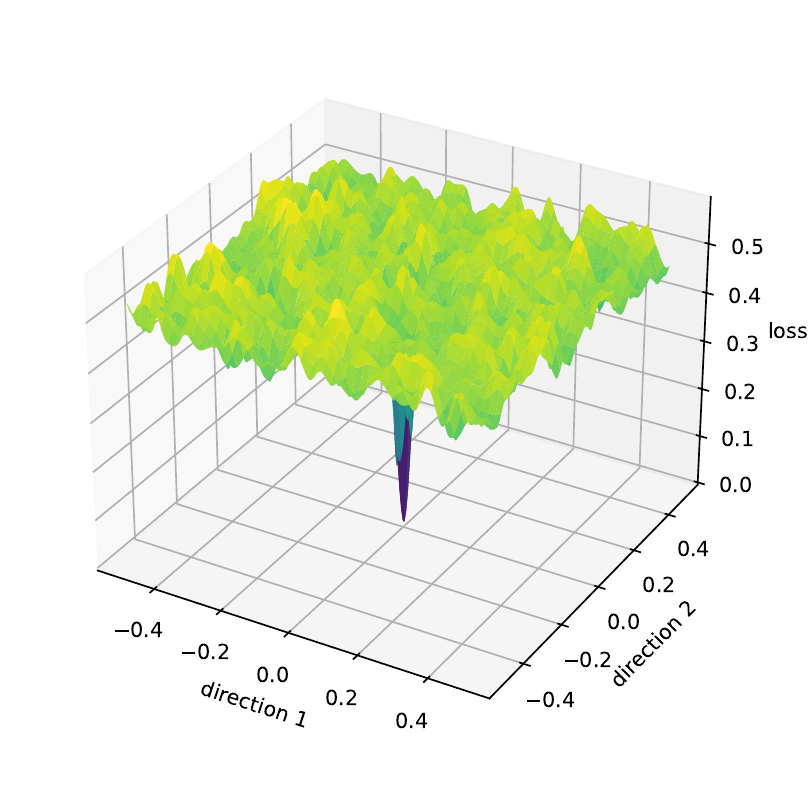}}
\subfigure{\includegraphics[width=0.23\textwidth]{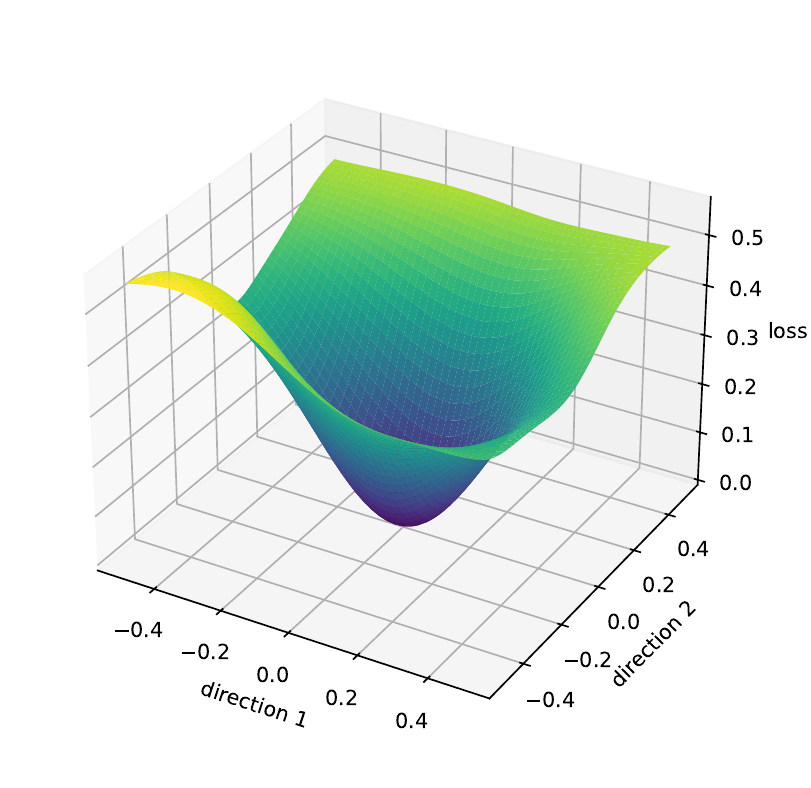}}
\subfigure{\includegraphics[width=0.23\textwidth]{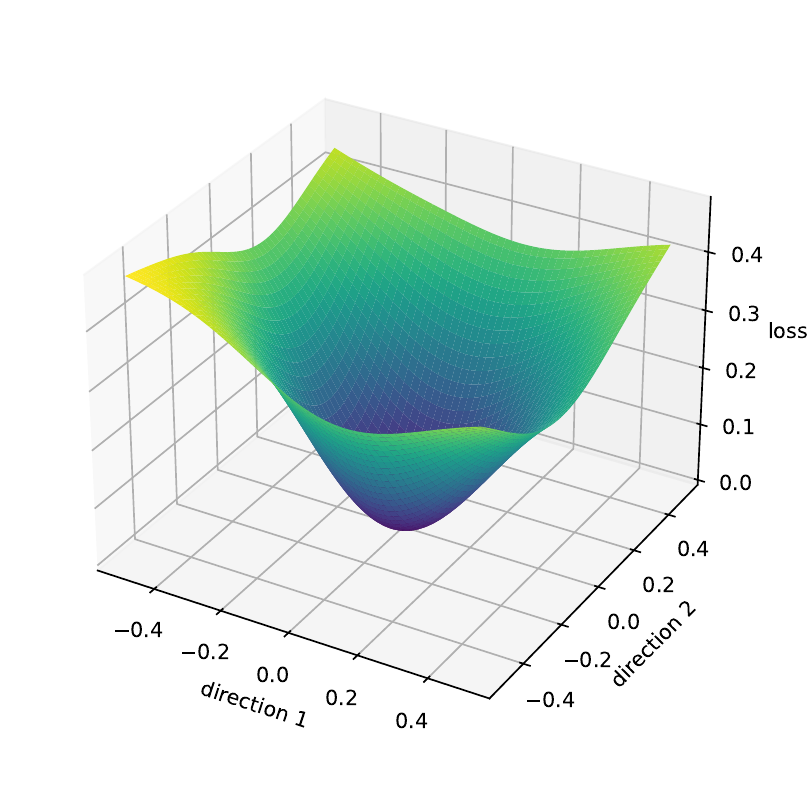}}\\
\vspace{-6mm}
\subfigure{\includegraphics[width=0.23\textwidth]{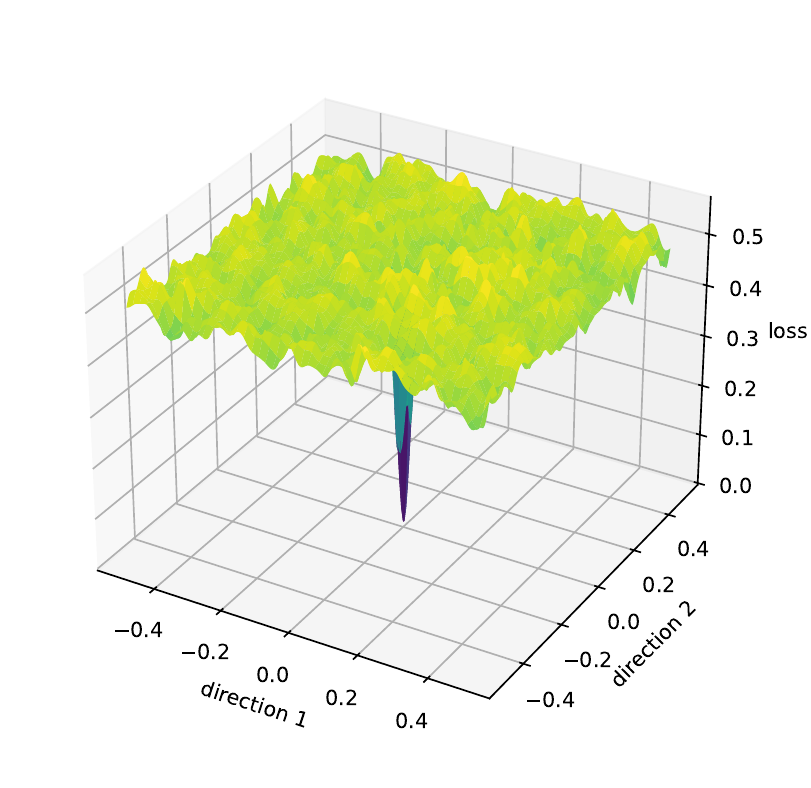}}
\subfigure{\includegraphics[width=0.23\textwidth]{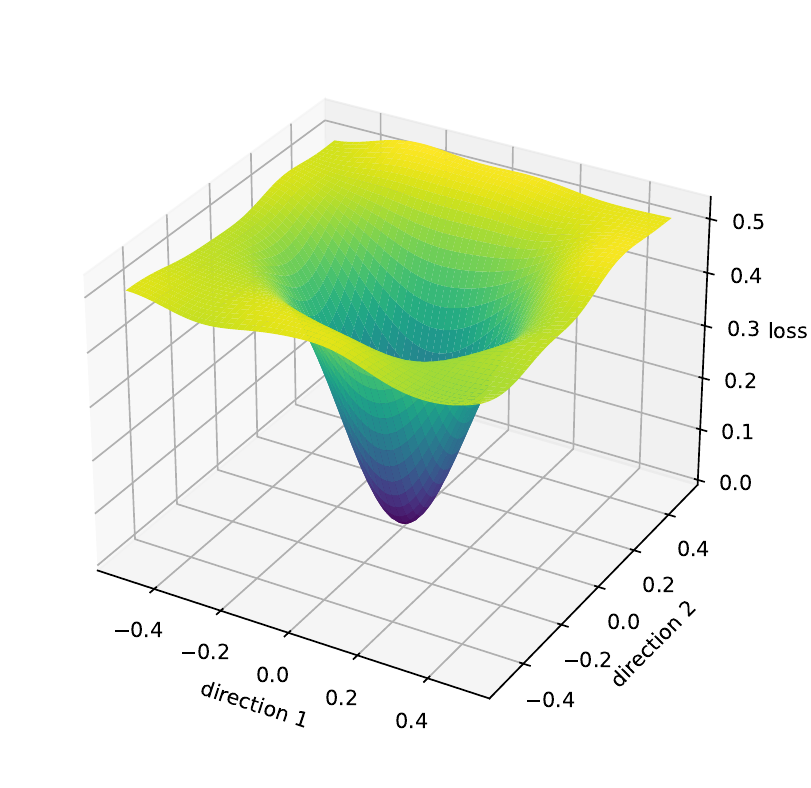}}
\subfigure{\includegraphics[width=0.23\textwidth]{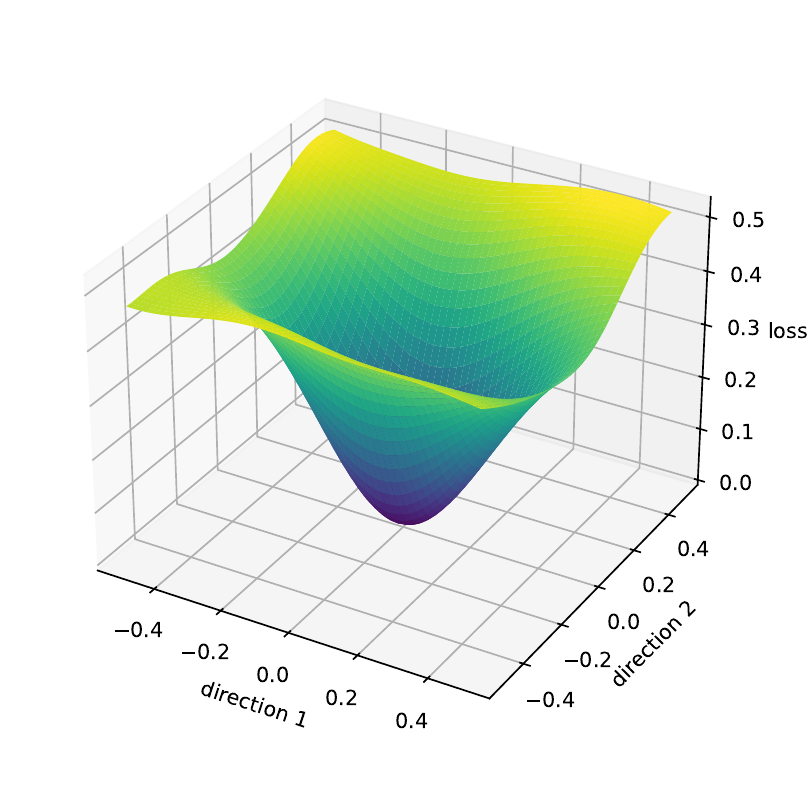}}\\
\vspace{-6mm}
\subfigure{\includegraphics[width=0.23\textwidth]{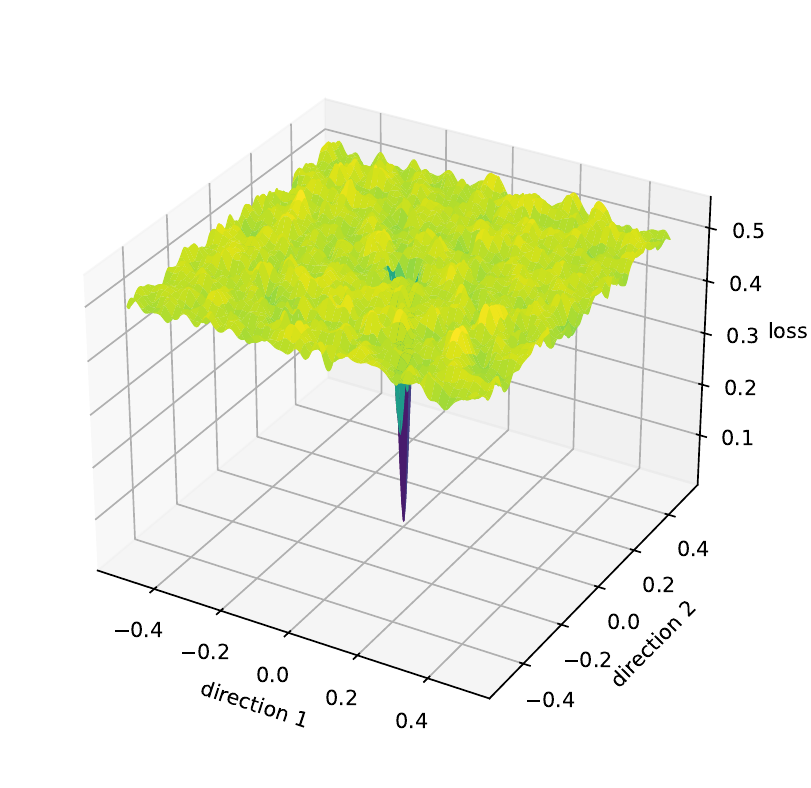}}
\subfigure{\includegraphics[width=0.23\textwidth]{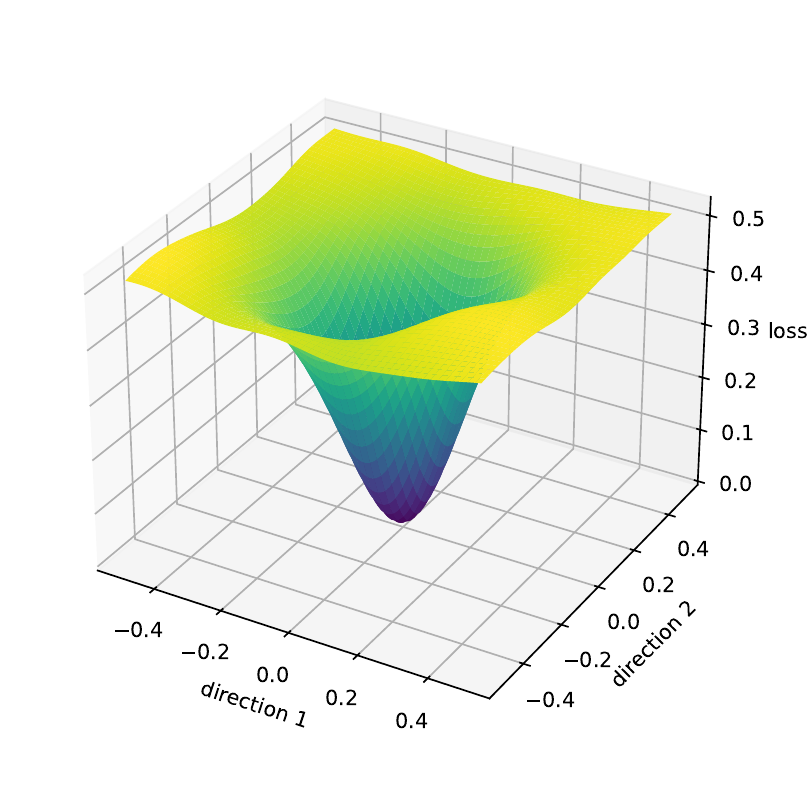}}
\subfigure{\includegraphics[width=0.23\textwidth]{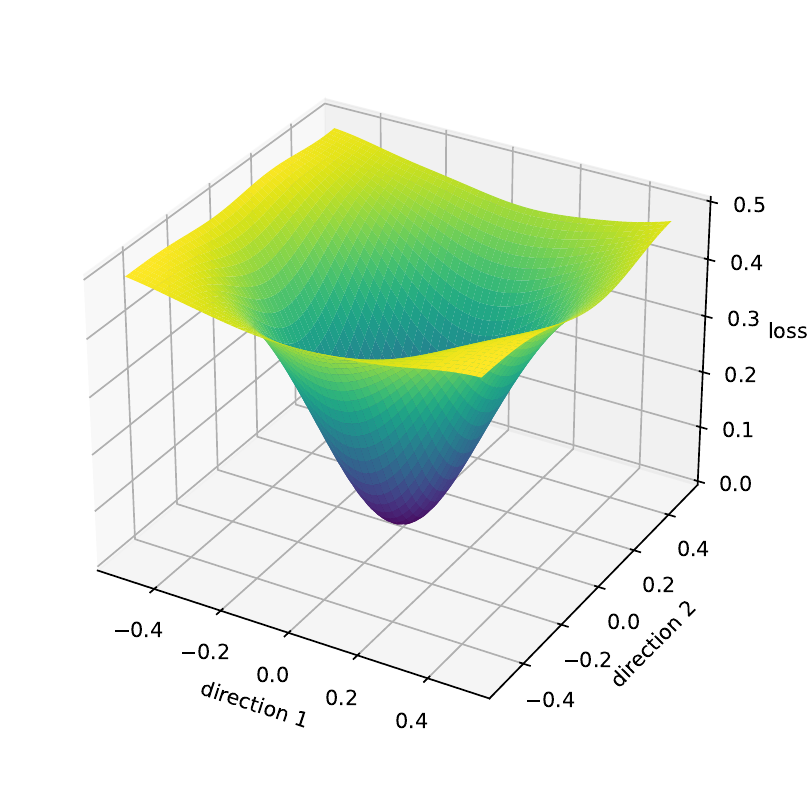}}\\

\caption{Loss landscapes for various  qubits counts and models. The left, middle, and right columns correspond to the SQC, NEQC-NN, and NEQC-CNN models, respectively. The rows, from top to bottom, correspond to 3 to 8 qubits.}
\label{loss_landscape}
\end{figure}

Figure \ref{loss_landscape} illustrates the loss landscapes for different qubit counts and models. For each qubit count and model, 10 runs are performed, resulting in 10 loss landscapes. Since these landscapes exhibit similar patterns, we show the landscape from the first run for each qubit count and model.
The landscape plots are arranged as follows: the left, middle, and right columns represent the SQC, NEQC-NN, and NEQC-CNN models, respectively. The rows, ordered from top to bottom, correspond to qubit counts ranging from 3 to 8. The results clearly show that the loss landscapes around the optimal solution for the NEQC-NN and NEQC-CNN models are smoother compared to those of the SQC model. Although the loss landscapes become less smooth as the number of qubits increases, the NEQC-NN and NEQC-CNN models consistently outperform the SQC model. For the SQC model, the number of local minima increases dramatically with a growing qubit count, and the global optimum becomes confined to a narrow gorge. However, this phenomenon is not observed in the NEQC-NN and NEQC-CNN models. These results highlight superior trainability of the NEQC-NN and NEQC-CNN models, particularly as the number of qubits increases.

Finally, we compare the expressibility of the SQC, NEQC-NN and NEQC-CNN models. As shown in Table \ref{expressibility}, the expressibility values of the NEQC-NN and NEQC-CNN model significantly surpasses those of the SQC model (a larger expressibility value means less expressive). This observation aligns with the relationship between expressibility and trainability discussed in Ref. \cite{erb}. By reducing expressibility, the NEQC-NN and NEQC-CNN model can decrease their susceptibility to BPs. These results explain the reason that using neural network to generate parameters can help mitigate BPs.

\begin{table}[!htbp]
    \centering
    \resizebox{0.9999\textwidth}{!}{
    \begin{threeparttable}
    \caption{Expressibility of circuits for varying qubit counts and model configurations}

    \begin{tabular}{lllllllllllll}
    \toprule
         qubit count& model & circuit 1 & circuit 2 & circuit 3 & circuit 4 & circuit 5 & circuit 6 & circuit 7 & circuit 8 & circuit 9 & circuit 10 & average \\
        \midrule
        \multirow{3}{*}{3} & \multirow{1}{*}{SQC} &
        $4.05\times10^{-3}$&
        $4.24\times10^{-3}$ &
        $3.71\times10^{-3}$ &
        $4.54\times10^{-3}$ &
        $4.16\times10^{-3}$ &
        $3.20\times10^{-3}$ &
        $3.46\times10^{-3}$ &
        $3.78\times10^{-3}$ &
        $4.53\times10^{-3}$ &
        $3.70\times10^{-3}$ &
        $3.94\times10^{-3}$
        \\

        ~ & \multirow{1}{*}{NEQC-NN}&
        \multirow{1}{*}{8.96} &
        \multirow{1}{*}{16.07} &
        \multirow{1}{*}{14.48} &
        \multirow{1}{*}{12.83} &
        \multirow{1}{*}{10.47} &
        \multirow{1}{*}{10.45} &
        \multirow{1}{*}{12.96} &
        \multirow{1}{*}{14.79} &
        \multirow{1}{*}{10.69} &
        \multirow{1}{*}{11.56} &
        \multirow{1}{*}{12.33}
        \\
        ~ & \multirow{1}{*}{NEQC-CNN}&
        \multirow{1}{*}{11.54} &
        \multirow{1}{*}{11.58} &
        \multirow{1}{*}{7.40} &
        \multirow{1}{*}{4.58} &
        \multirow{1}{*}{9.33} &
        \multirow{1}{*}{9.04} &
        \multirow{1}{*}{11.95} &
        \multirow{1}{*}{9.88} &
        \multirow{1}{*}{16.36} &
        \multirow{1}{*}{11.90} &
        \multirow{1}{*}{10.36}
        \\
        \\

        \multirow{3}{*}{4}  & \multirow{1}{*}{SQC} &
        $2.22\times10^{-3}$ &
        $3.24\times10^{-3}$ &
        $2.52\times10^{-3}$ &
        $2.50\times10^{-3}$ &
        $2.75\times10^{-3}$ &
        $2.41\times10^{-3}$ &
        $2.54\times10^{-3}$ &
        $1.95\times10^{-3}$ &
        $2.52\times10^{-3}$ &
        $1.89\times10^{-3}$ &
        $2.45\times10^{-3}$
        \\
        ~ & \multirow{1}{*}{NEQC-NN}&
        \multirow{1}{*}{12.49} &
        \multirow{1}{*}{11.24} &
        \multirow{1}{*}{21.84} &
        \multirow{1}{*}{23.67} &
        \multirow{1}{*}{26.54} &
        \multirow{1}{*}{22.13} &
        \multirow{1}{*}{14.20} &
        \multirow{1}{*}{25.11} &
        \multirow{1}{*}{16.46} &
        \multirow{1}{*}{17.53} &
        \multirow{1}{*}{19.12}
        \\
         ~ & \multirow{1}{*}{NEQC-CNN}&
        \multirow{1}{*}{6.02} &
        \multirow{1}{*}{10.37} &
        \multirow{1}{*}{13.91} &
        \multirow{1}{*}{13.87} &
        \multirow{1}{*}{6.41} &
        \multirow{1}{*}{9.70} &
        \multirow{1}{*}{23.49} &
        \multirow{1}{*}{13.88} &
        \multirow{1}{*}{5.21} &
        \multirow{1}{*}{17.83} &
        \multirow{1}{*}{12.07}
        \\
        \\

        \multirow{3}{*}{5}  & \multirow{1}{*}{SQC} &
        $1.81\times10^{-3}$ &
        $1.51\times10^{-3}$ &
        $9.62\times10^{-4}$ &
        $1.30\times10^{-3}$ &
        $9.08\times10^{-4}$ &
        $1.64\times10^{-3}$ &
        $1.60\times10^{-3}$ &
        $8.23\times10^{-4}$ &
        $1.19\times10^{-3}$ &
        $1.28\times10^{-3}$ &
        $1.30\times10^{-3}$
        \\
        ~ &
        \multirow{1}{*}{NEQC-NN}&
        \multirow{1}{*}{25.17} &
        \multirow{1}{*}{17.41} &
        \multirow{1}{*}{32.31} &
        \multirow{1}{*}{29.06} &
        \multirow{1}{*}{26.67} &
        \multirow{1}{*}{30.75} &
        \multirow{1}{*}{32.59} &
        \multirow{1}{*}{48.72} &
        \multirow{1}{*}{42.62} &
        \multirow{1}{*}{45.36} &
        \multirow{1}{*}{33.07}
        \\
        ~ & \multirow{1}{*}{NEQC-CNN}&
        \multirow{1}{*}{30.54} &
        \multirow{1}{*}{33.11} &
        \multirow{1}{*}{21.68} &
        \multirow{1}{*}{17.26} &
        \multirow{1}{*}{25.08} &
        \multirow{1}{*}{36.16} &
        \multirow{1}{*}{13.21} &
        \multirow{1}{*}{30.52} &
        \multirow{1}{*}{23.65} &
        \multirow{1}{*}{29.84} &
        \multirow{1}{*}{26.11}
        \\
        \\

        \multirow{3}{*}{6}  & \multirow{1}{*}{SQC} &
        $6.11\times10^{-4}$ &
        $6.33\times10^{-4}$ &
        $3.31\times10^{-4}$ &
        $9.61\times10^{-4}$ &
        $5.28\times10^{-4}$ &
        $1.24\times10^{-3}$ &
        $1.22\times10^{-3}$ &
        $8.49\times10^{-4}$ &
        $1.38\times10^{-3}$ &
        $5.19\times10^{-4}$ &
        $8.28\times10^{-4}$
        \\
        ~ &
        \multirow{1}{*}{NEQC-NN}&
        \multirow{1}{*}{60.00} &
        \multirow{1}{*}{82.01} &
        \multirow{1}{*}{56.24} &
        \multirow{1}{*}{90.51} &
        \multirow{1}{*}{47.33} &
        \multirow{1}{*}{32.48} &
        \multirow{1}{*}{62.92} &
        \multirow{1}{*}{67.79} &
        \multirow{1}{*}{26.64} &
        \multirow{1}{*}{51.99} &
        \multirow{1}{*}{57.79}
        \\
        ~ & \multirow{1}{*}{NEQC-CNN}&
        \multirow{1}{*}{86.80} &
        \multirow{1}{*}{16.97} &
        \multirow{1}{*}{33.06} &
        \multirow{1}{*}{14.90} &
        \multirow{1}{*}{45.91} &
        \multirow{1}{*}{34.75} &
        \multirow{1}{*}{11.02} &
        \multirow{1}{*}{64.52} &
        \multirow{1}{*}{13.85} &
        \multirow{1}{*}{39.51} &
        \multirow{1}{*}{36.13}
        \\
        \\

        \multirow{3}{*}{7}  & \multirow{1}{*}{SQC} &
        $3.45\times10^{-4}$ &
        $7.92\times10^{-4}$ &
        $3.71\times10^{-4}$ &
        $7.95\times10^{-4}$ &
        $7.98\times10^{-4}$ &
        $6.34\times10^{-4}$ &
        $3.05\times10^{-4}$ &
        $2.82\times10^{-4}$ &
        $2.44\times10^{-4}$ &
        $7.04\times10^{-4}$ &
        $5.27\times10^{-4}$
        \\
        ~ &
        \multirow{1}{*}{NEQC-NN}&
        \multirow{1}{*}{115.89} &
        \multirow{1}{*}{97.49} &
        \multirow{1}{*}{93.84} &
        \multirow{1}{*}{48.32} &
        \multirow{1}{*}{58.95} &
        \multirow{1}{*}{82.43} &
        \multirow{1}{*}{90.25} &
        \multirow{1}{*}{122.75} &
        \multirow{1}{*}{88.20} &
        \multirow{1}{*}{237.16} &
        \multirow{1}{*}{103.53}
        \\
       ~ & \multirow{1}{*}{NEQC-CNN}&
        \multirow{1}{*}{34.65} &
        \multirow{1}{*}{35.69} &
        \multirow{1}{*}{24.39} &
        \multirow{1}{*}{49.15} &
        \multirow{1}{*}{49.19} &
        \multirow{1}{*}{29.50} &
        \multirow{1}{*}{11.21} &
        \multirow{1}{*}{25.72} &
        \multirow{1}{*}{32.53} &
        \multirow{1}{*}{17.37} &
        \multirow{1}{*}{30.94}
        \\
        \\

        \multirow{3}{*}{8}  & \multirow{1}{*}{SQC} &
        $1.01\times10^{-4}$ &
        $1.16\times10^{-4}$ &
        $1.17\times10^{-4}$ &
        $2.10\times10^{-4}$ &
        $6.80\times10^{-5}$ &
        $6.64\times10^{-5}$ &
        $2.12\times10^{-4}$ &
        $7.13\times10^{-5}$ &
        $1.05\times10^{-4}$&
        $1.57\times10^{-4}$&
        $1.22\times10^{-4}$
        \\
        ~ &
        \multirow{1}{*}{NEQC-NN}&
        \multirow{1}{*}{Inf} &
        \multirow{1}{*}{Inf} &
        \multirow{1}{*}{Inf} &
        \multirow{1}{*}{Inf} &
        \multirow{1}{*}{Inf} &
        \multirow{1}{*}{Inf} &
        \multirow{1}{*}{Inf} &
        \multirow{1}{*}{Inf} &
        \multirow{1}{*}{Inf} &
        \multirow{1}{*}{Inf} &
        \multirow{1}{*}{Inf}
        \\
        ~ & \multirow{1}{*}{NEQC-CNN}&
        \multirow{1}{*}{Inf} &
        \multirow{1}{*}{Inf} &
        \multirow{1}{*}{Inf} &
        \multirow{1}{*}{Inf} &
        \multirow{1}{*}{Inf} &
        \multirow{1}{*}{Inf} &
        \multirow{1}{*}{Inf} &
        \multirow{1}{*}{Inf} &
        \multirow{1}{*}{Inf} &
        \multirow{1}{*}{Inf} &
        \multirow{1}{*}{Inf}
        \\
        \bottomrule
    \end{tabular}

\label{expressibility}
\end{threeparttable}
}
\end{table}

\section{Conclusion}
\label{Sect:conclusion}
In this research, we enhance the effectiveness and broaden the applicability of neural networks in mitigating the BP problem during the training of VQCs. We optimize the neural network architecture, extend the circuit input from classical data encoding to random quantum states, and generalize the VQC structure from a fixed to a random configuration. These modifications enhance the NN-based BP mitigation approach, making it more applicable to VQAs designed to prepare a target quantum state, such as QAOA and VQE, as well as to VQAs with flexible circuit structures, like those in quantum architecture search.

We investigate the NN-based BP mitigation approach from several aspects, including the number of iterations required for convergence, the smoothness of the loss landscape, and expressibility. Our results demonstrate that employing neural networks significantly reduces the number of iterations required for the VQC to converge to the global optimum. Visualization of the loss landscapes reveal that the neural network approach yields a smoother landscape, making it less susceptible to BPs, whereas the original VQC displays a rugged landscape with the global optimum lying in a very narrow gorge. Additionally, the NN-enhanced model exhibits limited expressibility, which, according to the established relationship between expressibility and trainability, contributes to improved trainability. Our research demonstrates that the NN-based BP mitigation approach possesses a degree of universality, making it valuable for advancing the development of VQAs across various fields.

\section*{Acknowledgements}

This work is supported by National Natural Science Foundation of China (No. 62471187),
Guangdong Basic and Applied Basic Research Foundation (Nos. 2022A1515140116, 2021A1515011985),
and Innovation Program for Quantum Science and Technology (No. 2021ZD0302901).

\end{document}